\documentclass{aastex6}
\usepackage{mathtools}
\usepackage{amsmath}
\usepackage{float}
\usepackage{longtable}
\usepackage{soul}
\DeclareMathOperator{\sinc}{sinc}
\fullcollaborationName{The Friends of AASTeX Collaboration}

\begin{document}


\title{ARMADA II: Further Detections of Inner Companions to Intermediate Mass Binaries with Micro-Arcsecond Astrometry at CHARA and VLTI}


\author{Tyler Gardner\altaffilmark{1}, John D. Monnier\altaffilmark{1}, Francis C. Fekel\altaffilmark{2}, Jean-Baptiste Le Bouquin\altaffilmark{3}, Adam Scovera\altaffilmark{1}, Gail Schaefer\altaffilmark{4}, Stefan Kraus\altaffilmark{5}, Fred C. Adams\altaffilmark{1,6}, Narsireddy Anugu\altaffilmark{4}, Jean-Philippe Berger\altaffilmark{3}, Theo Ten Brummelaar\altaffilmark{4}, Claire L. Davies\altaffilmark{5}, Jacob Ennis\altaffilmark{1}, Douglas R. Gies\altaffilmark{4}, Keith J.C. Johnson\altaffilmark{7}, Pierre Kervella\altaffilmark{8}, Kaitlin M. Kratter\altaffilmark{9}, Aaron Labdon\altaffilmark{10}, Cyprien Lanthermann\altaffilmark{4}, Johannes Sahlmann\altaffilmark{11}, Benjamin R. Setterholm\altaffilmark{1}}

\altaffiltext{1}{Astronomy Department, University of Michigan, Ann Arbor, MI 48109, USA}
\altaffiltext{2}{Center of Excellence in Information Systems, Tennessee State University, Nashville, TN 37209, USA}
\altaffiltext{3}{Univ. Grenoble Alpes, CNRS, IPAG, 38000 Grenoble, France}
\altaffiltext{4}{The CHARA Array of Georgia State University, Mount Wilson Observatory, Mount Wilson, CA 91203, USA}
\altaffiltext{5}{Astrophysics Group, Department of Physics \& Astronomy, University of Exeter, Stocker Road, Exeter, EX4 4QL, UK}
\altaffiltext{6}{Physics Department, University of Michigan, Ann Arbor, MI 48109, USA}
\altaffiltext{7}{Department of Computer Sciences, University of Wisconsin, Madison, WI 53706, USA}
\altaffiltext{8}{LESIA, Observatoire de Paris, Universit\'e PSL, CNRS, Sorbonne Universit\'e, Universit\'e Paris Cit\'e, 5 place Jules Janssen, 92195 Meudon, France}
\altaffiltext{9}{Department of Astronomy and Steward Observatory, Univ. of Arizona, 933 N Cherry Ave, Tucson, AZ, 85721, USA}
\altaffiltext{10}{European Southern Observatory, Casilla 19001, Santiago 19, Chile}
\altaffiltext{11}{RHEA Group for the European Space Agency (ESA), European Space Astronomy Centre (ESAC),\\ Camino Bajo del Castillo s/n, 28692 Villanueva de la Ca\~nada, Madrid, Spain}

\begin{abstract}
We started a survey with CHARA/MIRC-X and VLTI/GRAVITY to search for low mass companions orbiting individual components of intermediate mass binary systems. With the incredible precision of these instruments, we can detect astrometric ``wobbles" from companions down to a few tens of micro-arcseconds. This allows us to detect any previously unseen triple systems in our list of binaries. We present the orbits of 12 companions around early F to B-type binaries, 9 of which are new detections and 3 of which are first astrometric detections of known RV companions. The masses of these newly detected components range from 0.45--1.3 M$_{\odot}$. Our orbits constrain these systems to a high astrometric precision, with median residuals to the orbital fit of 20-50$\mu$as in most cases. For 7 of these systems we include newly obtained radial velocity data, which help us to identify the system configuration and to solve for masses of individual components in some cases. Although additional RV measurements are needed to break degeneracy in the mutual inclination, we find that the majority of these inner triples are not well-aligned with the wide binary orbit. This hints that higher mass triples are more misaligned compared to solar and lower mass triples, though a thorough study of survey biases is needed. We show that the ARMADA survey is extremely successful at uncovering previously unseen companions in binaries. This method will be used in upcoming papers to constrain companion demographics in intermediate mass binary systems down to the planetary mass regime.  
\end{abstract}

\keywords{astrometry, binaries: close, technique: interferometry}



\section{Introduction} 
\label{sec:intro}
Binary stars provide the unique opportunity to measure the stellar masses of the components. If astrometric measurements are combined with radial velocity (RV) data, one can unambiguously measure individual masses and physical orbital elements of the system \citep[e.g.][]{bonneau2014}. The observation of binary systems is one of the few ways to obtain such direct mass measurements, making their study a continued crucial aspect of the field for comparison with stellar evolution models. However, precision measurements of mass require a good knowledge of the binary orbit. Well-constrained orbits are hard to obtain for long period binary systems, since we often only observe a small arc of their visual orbit and the RVs of visual binary components are often slowly varying with little RV difference and have low amplitudes making them difficult to measure. Combining data over many decades is necessary to obtain precision orbits for binaries with orbital periods of tens to hundreds of years. 

In addition to their importance for mass determinations, binary and multiple star systems are also valuable as testbeds for studying formation mechanisms and in
expanding multiplicity statistics. As our instruments and methods continue to improve, we are now able to begin measuring astrometric binary orbits at the few tens of micro-arcsecond ($\mu$as) level of precision. With such high precision, one can search for the astrometric ``wobble'' due to additional unseen companions in the system down to the planetary mass regime (e.g. see \citealt{Muterspaugh2010,gardner2018,gardner2021}). This method opens up new regimes to hunt for previously unseen companions orbiting individual components of long period binary systems. Such systems are often hard to detect with other methods, particularly for intermediate mass stars where RV measurements are difficult due to the weak and broad spectral lines of such stars. Hence, by detecting and characterizing new companions in these regimes we can better understand the multiplicity and formation mechanisms of such compact multiple systems. 

We started the ARrangement for Micro-Arcsecond Differential Astrometry (ARMADA), a survey that uses ground-based long baseline optical interferometers for the purpose of detecting previously unseen companions in intermediate mass binary systems down to the planetary mass regime. Our observational methods, an astrometric test source, and first new detections are described in detail in \citet{gardner2021}. For the ARMADA survey we target late B-type, A-type, and early F-type systems where low mass companion frequency in the $\sim$au regime is difficult to measure with other methods. For the exoplanets, this regime is relatively unprobed due to the weak and broad spectral lines from the rapidly rotating primary star. Though direct imaging has been successful at discovering wide-orbit planets around isolated A-type stars \citep{nielsen2019,vigan2021}, the $\sim$au regime is too close-in to probe with this method. The multiplicity rate of inner stellar (triple) companions to outer binaries for intermediate mass systems is also somewhat uncertain. For intermediate mass stars in general, multiplicity surveys are rather incomplete for inner companions with M$_{\rm{companion}}$ / M$_{\rm{star}}$ $<$ 0.4 in 0.5-5 au orbits (see Fig. 1 of \citealt{moe2017}; De Furio et al, in prep). These statistics are even more uncertain for the inner triples of 10-100 au binaries, since the observational methods become more difficult. We know that in general triple systems appear to be more common for higher mass stars \citep{maz2019}, making high mass stars a potentially ripe regime for detecting inner subsystems to binaries. Our first discoveries are compact triple systems, since they impart a larger ``wobble'' onto their host star compared to a planet. This allows us to build up detections to improve the multiplicity statistics in these regimes, as well as to better understand formation mechanisms and orbital configurations of such compact triple systems. 

Optical long-baseline interferometry is an important method for obtaining precision orbits and masses of visual binaries, due to the high angular resolution and astrometric precision possible with such facilities. Current instruments are able to obtain astrometric measurements of close binaries down to the $\sim$10$\mu$as precision level (e.g., GRAVITY at VLTI: \citealt{gravity2017}; MIRC/MIRC-X at CHARA: \citealt{gardner2018, gardner2021}, \citealt{schaefer2016}). In this paper we present data taken for the continuation of the ARMADA survey with the MIRC-X instrument \citep{anugu2020}, which is a recent upgrade of the Michigan Infra-Red Combiner (MIRC) \citep{Monnier2006}. MIRC-X is a $H$-band combiner of six 1~m telescopes at the Georgia State University Center for High Angular Resolution Astronomy (CHARA) Array. We also include data taken from the GRAVITY instrument at the Very Large Telescope Interferometer (VLTI), which combines the four 1.8-meter Auxiliary Telescopes (ATs) in the $K$-band \citep{gravity2017}. In this paper we present 12 compact triple orbits detected with ARMADA data, 3 previously known triples for which we are able to improve the orbit, and 9 newly detected systems. To help determine the configuration of the systems and solve for the masses and mutual inclinations of these systems, we include new RV data taken with the Tennessee State University 2-meter Automated Spectroscopic Telescope (AST) at the Fairborn Observatory \citep{ew07}. However, we note that some of the A/B-type systems are too broad-lined to be able to measure the RV variations and our RV survey is ongoing.

In section \ref{sec:observations} we describe our observations and data reduction methods for ARMADA. Section \ref{sec:orbitfitting} outlines our orbit fitting models. Section \ref{sec:detections} shows our orbital fits to 12 systems, including systems with RV data, systems where we detect the flux of the inner component, and systems with ``wobble"-only orbits. We also include a brief literature review and notes for individual systems in this section. We discuss the implications of high mutual inclinations detected in these systems in Section \ref{sec:inclinations}. We give concluding remarks and prospects for future results in section \ref{sec:conclusion}.

\section{Observations and Data Reduction} 
\label{sec:observations}
\subsection{MIRC-X at the CHARA Array}
Data for the ARMADA survey are taken in $H$-band with the Michigan InfraRed Combiner-Exeter (MIRC-X) instrument at the CHARA Array. The CHARA Array is the optical/near-IR interferometer with the longest baselines in the world \citep{Brummelaar2005}. MIRC-X combines all six telescopes available at CHARA with baselines up to 330 m. The original MIRC instrument is described in detail by \citet{Monnier2006}, and in 2017 July the detector and combiner were upgraded to MIRC-X \citep{anugu2020}. Observations for the ARMADA survey are all taken in grism (R$\sim$190) mode, allowing us to detect components out to $\sim$200 mas with the larger interferometric field-of-view. 

The MIRC-X combiner measures visibilities, differential phases, and closure phases of our targets. Normally, one employs frequent observations of nearby calibrator stars to measure visibility loss due to time-variable factors such as atmospheric coherence time, vibrations, differential dispersion, and birefringence in the beam train. For ARMADA our main interest is differential astrometry between two components of a binary system within the interferometric field-of-view (both components unresolved). Since closure phase is immune to atmospheric effects, and extra dispersion in differential phase can be fit with a polynomial, we are able to observe for ARMADA without the use of traditional calibrators by fitting to closure and differential phase as described in \citet{gardner2021}. Our observational setup and calibration methods follow the procedures outlined in that paper. We used the MIRC-X data pipeline (version 1.3.3) to produce OIFITS files for each night, described in \citet{anugu2020}. This pipeline and its documentation is maintained on Gitlab\footnote{\url{https://gitlab.chara.gsu.edu/lebouquj/mircx_pipeline}}. These data were reduced with the ``spectral-differential'' method of the MIRC-X pipeline for computing differential phase. This method first removes the group delay from the raw phase, and then computes differential phase as the phase(i+1) - phase(i) where i, i+1 are neighboring wavelength channels. We reduced our data with the number of coherent integration frames (ncoh) of 10, a maximum integration time of 60 seconds for each measurement of visibility and phase within the oifits files, and bispectrum bias correction applied.

The nominal wavelength knowledge precision for MIRC/MIRC-X has been shown to be at the $\sim10^{-3}$ level \citep{monnier2012}. For a 100 milli-arcsecond (mas) binary, this would imply a potential systematic error at the 100 $\mu$as level when combining epochs of astrometry data. To do better than this, we employ a special calibration mode which uses etalons of different thicknesses which simulate the signal of a binary system. We take data with these etalons at least once per night on the CHARA Array's Six Telescope Simulator (STS) lab source. Briefly, since the thicknesses of the etalons are known to a high precision, we can solve for an ``astrometric correction factor'' which brings all CHARA ARMADA nights to the same wavelength scale. This mitigates the systematic uncertainties between epochs due to wavelength calibration, and brings our relative systematic errors closer to the $10^{-4}$ level (for a detailed description of the etalon calibration scheme, see \citealt{gardner2021}).  

\subsection{GRAVITY at VLTI}
To cover southern sky targets for the ARMADA survey, we use data from the GRAVITY instrument at VLTI \citep{gravity2017}. ARMADA data are taken in service mode under the large program ID 1103.C-0477. For our service-mode program, GRAVITY combines the four 1.5-meter Auxiliary Telescopes (ATs) in the K band with baselines up to 130~m. We utilize the large, astrometric, and medium configurations to best constrain the astrometry of our wide ($\sim$100-200 mas) binaries. Our data are recorded in GRAVITY's high resolution mode, with R$\sim$4000. This high spectral resolution is needed to achieve our desired wavelength knowledge for precision astrometry. Our data are taken with a Detector Integration Time (DIT) of 30 seconds, which prevents signal loss due to the fast-varying phase and visibility changes of wide binary stars. Each pointing has 8 frames (NDIT = 8), with an observing sequence OOSOOOSOOO (O=object, S=sky). With this sequence a single epoch on an ARMADA binary is completed in just under an hour when including overhead time. To make our large number of target observations feasible, we do not use calibrator stars (just as for the MIRC-X/CHARA data). Hence, our binary astrometry is again largely set by the variation of signal with wavelength for the closure phase. 

We used the standard reduction pipeline for GRAVITY data\footnote{\url{https://www.eso.org/sci/software/pipelines/gravity/}}, version 1.5.0, along with the suite of Python tools developed by the GRAVITY team for running the pipeline. When reducing the data, we set the maximum number of frames to 2, which means each measurement of phase and visibility is $\leq$60 seconds of integration. Since the wavelength variation of visibility and phase sets the binary astrometry, we are able to fit directly to the uncalibrated closure phase. For GRAVITY data, we see very minimal loss in the visibility amplitude (VISAMP) over the course of an observation. Thus, we also find VISAMP to be a useful observable for the GRAVITY/VLTI data. Note that the GRAVITY instrument uses a Fourier transform spectrometer source to measure absolute wavelength scales at the $5 \times 10^{-5}$ level \citep{sanchez2017}. For a binary with 200 mas separation, this corresponds to 20$\mu$as astrometric precision. Hence, we do not need to employ an extra wavelength calibration step for each ARMADA night as for MIRC-X. For shared ARMADA sources between GRAVITY/MIRC-X instruments, we can then bring our MIRC-X data to the same absolute wavelength scale as described for sources HD 199766, HD 29573, and HD 31297.

\subsection{Fitting Binary Star Differential Astrometry}
We use the fitting tools described in \citet{gardner2021} to fit a binary model to the interferometry observables. The free parameters for this binary model include a uniform disk for the primary and secondary to form visibilities $V_1$ and $V_2$; a binary separation in right ascension (R.A.) and declination (DEC) -- ($\alpha$, $\delta$); a monochromatic flux ratio between the two components $f$; as well as a bandwidth smearing parameter $b = 1/R$, where $R$ is the resolution of the disperser and $\Gamma = \sinc[b(u\alpha+v\delta)]$. The location on the uv-plane is denoted by the spatial frequencies $u$ and $v$.

Since we do not use the standard CAL-SCI sequence of observing for MIRC-X or GRAVITY, our squared visibilities are poorly calibrated. For MIRC-X we thus use the closure phase and differential phase observables to fit our binary positions for each epoch. However, GRAVITY references its differential phase measurement to the low resolution fringe tracker instrument (which itself is pointed to the same source, in our case a binary with variable phase with wavelength). Hence we do not include differential phase in our binary fits for GRAVITY, although for GRAVITY data we do include visibility amplitude (VISAMP) in our fits. The VISAMP does not show significant loss due to lack of calibration, and it is the variation with wavelength again which largely sets our binary astrometry. For VISAMP, we include in our models free parameters for a third-order polynomial for each baseline (this captures non-source signal that would normally be calibrated out with on-sky calibrator sources). We fix the uniform diameter (UD) values in our fits to 0.5 mas, since these targets are all unresolved. This choice has no effect on the fitted astrometry within our measurement error bars.

As described in \citet{gardner2021}, to find our best differential astrometry solution on a given night we first perform a wide grid search in R.A. and DEC with step sizes of 0.1 milli-arcseconds to find the minimum $\chi^2$ solution. We then use the $lmfit$ package in Python to narrow in on the best solution with a non-linear least squares fit \citep{newville2016}. We convert our astrometry solutions from differential R.A. and DEC to a separation and position angle (PA) east of north pointing from the primary to secondary ($\rho$, $\theta$). After finding the best-fit $\rho$ and $\theta$, we need to compute the errors in position. In general it is very difficult to accurately estimate the astrometric errors of individual data points. One method involves mapping the 2D surface where the raw $\chi^2$ increases by 1 or 2.3 from the minimum value \citep{press1992}. However, in practice this often leads to errors which are underestimated. To avoid this underestimation, we start with a conservative case to accurately capture the shape of our astrometric error. We map out the 2D surface in ($\Delta$R.A., $\Delta$DEC) where the reduced $\chi^2$ increases by 1 from the minimum value (i.e. conservatively assuming fully correlated error bars that cannot be averaged). Often one uses the raw chi2 to estimate error ellipses in order to account for relative error sizes between epochs. In the case of ARMADA, however, our observational setup was uniform across all epochs, meaning that our degrees of freedom do not change significantly. In this special case, using the reduced chi2 statistic to estimate error ellipse sizes is sufficient. This leads to a positional error ellipse of accurate shape, but with major and minor axes sizes that are generally overestimated. We later scale the size of these errors when performing our full set of orbital fits described in section \ref{sec:orbitfitting}, so that each independent dataset contributes a final orbital fit $\chi^2_{red}=1$. We report the binary star astrometry and final astrometric errors for each object in Table \ref{table:astrometry}.

\begin{table}[h]
\centering
\caption{Binary Star Astrometry\tablenotemark{a}}
\label{table:astrometry}
\begin{tabular}{lccccccc}
\hline
\colhead{HD \#} & \colhead{MJD} & \colhead{Sep (mas)} & \colhead{PA (deg)\tablenotemark{b}}   & \colhead{Err Major (mas)} & \colhead{Err Minor (mas)} & \colhead{Err PA (deg)\tablenotemark{c}} & Instrument \\
\hline
199766 (A,B) & 58637.4729 & 109.130 & 279.397 & 0.027 & 0.012 & 341.448 & MIRC-X \\
& 58668.3154 & 103.820 & 279.198 & 0.042 & 0.019 & 348.616 & GRAVITY \\
& 58691.2737 & 100.3129 & 279.0101 & 0.1049 & 0.0507 & 342.6587 & GRAVITY \\
& 58695.3083 & 100.155 & 278.963 & 0.03 & 0.0181 & 323.823 & MIRC-X \\
& 58696.2998 & 99.937 & 278.898 & 0.022 & 0.0124 & 347.503 & MIRC-X \\
& ... & ... & ... & ... & ... & ... \\
\hline
1976 (A,B) & 58702.4063 & 113.286 & 137.451 & 0.157 & 0.135 & 129.749 & MIRC-X \\
& 59044.4081 & 114.596 & 140.422 & 0.093 & 0.055 & 307.211 & MIRC-X \\
& 59152.3097 & 116.075 & 141.388 & 0.083 & 0.064 & 114.084 & MIRC-X \\
& 59153.2224 & 115.925 & 141.396 & 0.121 & 0.073 & 102.523 & MIRC-X \\
... & ... & ... & ... & ... & ... & ... \\
\hline
\end{tabular}
\tablenotetext{a}{Full table available online}
\tablenotetext{b}{Postion angle E of N pointing from primary (brightest in H/K) to secondary}
\tablenotetext{c}{Position angle E of N of the error ellipse major axis}
\end{table}

\subsection{Spectroscopic Observations and Reductions with the TSU 2-m AST}
From 2021 January through 2022 March we obtained
spectra of our suspected triple systems at Fairborn
Observatory in southeast Arizona \citep{ew04}. We 
acquired the observations with the Tennessee State 
University 2~m Automatic Spectroscopic Telescope (AST), 
a fiber-fed echelle spectrograph \citep{ew07}, and a
Fairchild 486 CCD that has a 4K $\times$ 4K array of 15 
$\mu$m pixels \citep{fetal13}.  The size of the array 
results in a wavelength coverage that ranges from 3800~\AA~ 
to 8260~\AA. The spectra have a resolution of 0.24~\AA,
corresponding to a resolving power of 25000 at 6000~\AA. 
The best spectra have signal-to-noise ratios of about 150.

\citet{ftw09} have provided a general description of
the typical velocity reduction, which for solar-type 
stars is done in the spectral region 
4920\AA--7100~\AA. Given the broader range of
spectral types for the ARMADA project, several different 
line lists were used. For the F-type stars and those 
binaries with composite spectra that include a G/K giant,
we obtained velocities with our solar-type star line 
list, which consists of mostly neutral iron lines. 
For the binaries with
A-type primaries, we initially examined their spectra
with our A star line list, which contains ionized metal
lines and covers the above mentioned wavelength region.
Lines of the A- and early F-type components are 
often detectable with both our A-star and solar line 
lists, but the solar list with its much larger number 
of lines generally produces an average profile with 
higher signal-to-noise ratio. The cooler visual 
secondaries were also usually more easily detected with
the solar-type line list. To fit the individual 
line profiles we used a rotational broadening function
\citep{fg11, lf11}. For several systems two sets 
of lines, typically with very different line widths
and strengths, were detected. In the cases where both
sets of lines are always at least partially blended,
the line profiles of both components were fitted
simultaneously with two rotational broadening functions.

In a small number of cases we also used two other 
line lists that are blueward of our usually used
4920~\AA--7100~\AA~region. A list of 31 lines of
singly ionized metals situated between 4400~\AA~and
4920~\AA~was used for spectral classes around A0
and for A stars with $v$~sin~$i$ $>$100 km~s$^{-1}$.
For the B-type stars a very different line list was
required. We measured velocities of those systems with 
a line list that included six He~I lines between 4300~\AA~
and 5100~\AA~plus the Mg~II line at 4481.224~\AA~and the 
Si~II line at 5055.949~\AA.

Unpublished velocities that have been obtained with the
AST, its echelle spectrograph, and the Fairchild 486 
CCD for several IAU solar-type velocity standard stars
show that our velocities with the Fairchild CCD have a
$-$0.6 km~s$^{-1}$ shift relative to the results of
\citet{s10}, so we have added 0.6~km~s$^{-1}$ to our
velocities obtained with both the A and solar lines
lists that cover the same 4920~\AA--7100~\AA~region. 
For A stars measured with the A star list that
covers the blue wavelengths and for the B-type stars, 
no zero-point correction has been made to their velocities.

The rotational broadening fits enable us to determine
the projected rotational broadening of many of the 
components. For $v$~sin~$i$ values greater than 100 
km~s$^{-1}$ the estimated uncertainty is 4 km~s$^{-1}$,
Above 60~km~s$^{-1}$ the uncertainty is 2--3 km~s$^{-1}$
while below that value it is 1 km~s$^{-1}$. Table \ref{table:rv} 
reports our previously unpublished measurements of RV. 

\begin{table}[H]
\centering
\caption{Binary Star RV\tablenotemark{a}}
\label{table:rv}
\begin{tabular}{lccc}
\hline
\colhead{HD \#} & \colhead{MJD\tablenotemark{b}} & \colhead{RV (km s$^{-1}$)} & \colhead{RV Err (km s$^{-1}$)} \\
\hline
199766 (Aa) & 59296.5142 & -27.2 & 2 \\
& 59309.4629 & 35.3 & 2 \\
& 59315.4621 & 31.4 & 2 \\
& 59317.4626 & 28.6 & 2 \\
& 59318.4628 & -25.9 & 2 \\
... & ... & ... \\
\hline

\hline
\end{tabular}
\tablenotetext{a}{Full table available in online format}
\tablenotetext{b}{Modified Julian Date}
\end{table}

\section{Orbit Fitting}
\label{sec:orbitfitting}
Once we measure our binary positions for each night, we perform a Keplerian orbit fit to our data with our Python routines described in \citet{gardner2021}. The Campbell elements ($\omega$, $\Omega$, $e$, $i$, $a$, $T$, $P$) describe the Keplerian motion of the brighter star of a binary system relative to the other, and relative to the observer. Those symbols have their usual meanings where $\omega$ is the longitude of the periastron, $\Omega$ is the position angle of the ascending node, $e$ is the eccentricity, $i$ is the orbital inclination, $a$ is angular separation, $T$ is a time of periastron passage, and $P$ is the orbital period. When including RV data, we also fit the semi-amplitudes $K$ and systemic velocity $\gamma$. The longitude of periastron $\omega$ is traditionally reported for the secondary when fitting to visual binary orbits alone. The convention when combining RV orbits is to report $\omega$ of the primary, which is flipped by 180$^{\circ}$. When we have RV data included in the orbit, we report the $\omega$ of the primary for these orbits. For purely visual orbits, there is a 180 degree ambiguity between $\omega$ and $\Omega$. In these cases we report the case where $\Omega < 180^{\circ}$. When performing fits, we convert these parameters to the linear Thieles-Innes coefficients as described in \citet{Wright2009}. We again use the $lmfit$ package in Python for non-linear least squares fitting.

When fitting a system of three or more components for ARMADA, we assume the system is hierarchical with the wide companion orbiting the center-of-mass of the inner pairs. This means that our orbit model is simply a sum of the outer plus inner Keplerian orbits. Most of the outer orbits we present in this paper are significantly larger than the inner orbits ($>$200 times larger in orbital period, with the exceptions of HD 220278 and HD 185762), so this hierarchical model is a reasonable assumption. Our inner orbital elements are then describing the ``wobble'' motion of one star about the center-of-mass of the inner orbit. In this case, the angular semi-major axis $a_{wob}$ of the tertiary component describes the size of the wobble motion, where one would need to know the mass ratio or detect the component directly in flux to figure out the true angular semi-major axis of the inner pair. Note that for unresolved inner binaries, the ``wobble" component measures the center-of-light position of the inner orbit. If the secondary flux is negligible, this position matches the motion of the primary around the center-of-mass of the inner orbit. Otherwise the center-of-light motion will be smaller. 

Our systems have long orbital periods, meaning that ARMADA data alone cannot constrain the outer binary orbit. Though we can still search for inner companions when there are degeneracies on the outer orbital elements, we include historical data from the Washington Double Star (WDS) catalog to better constrain these outer orbits \citep{mason2001}. Since we have high-precision differential astrometry, we correct for the precession of north when combining position angles measured by MIRC-X/GRAVITY to historical data in the WDS catalog (described in \citealt{gardner2021}). Errors on binary positions are not well known in this catalog, since it compiles data taken from many different surveys and methods. Based on the large scatter of these data about their best fits, we first assign circular errors of radius 10 milli-arcseconds to the WDS data taken with speckle methods and 50 milli-arcsecond errors for all other observational methods. We later adjust these errors so that the ARMADA set and WDS set are both contributing a reduced $\chi^2$ of 1. We use the ORB6\footnote{\url{http://www.astro.gsu.edu/wds/orb6.html}} catalog for initial guesses of the outer pair's orbital parameters. Once we find the best fit for the outer binary, we begin searching for the inner companion. To do so, we vary the inner orbital period on a grid and fit circular inner orbits to each fixed period \citep{gardner2021}. The outer binary elements are also varied at each step as free parameters, with the initial best-fit outer binary used as a starting point. Our period search grid follows that of \citet{muterspaugh_limits}, which is inspired by Nyquist frequency sampling. For data spanning a time $T$, we search periods $P = 2fT / k$ where $k$ is an integer and $f$ an oversampling factor set to 3 to ensure we do not miss companions due to uneven sampling. The minimum astrometric period that we evaluate is 2 days, so this sets the maximum value for $k$. Once the best inner period is detected, we refine our search further by performing a joint outer + inner fit using the parameter guesses we found in the previous method as starting values.

To compute error bars on our parameters, we determine posterior distributions on our orbital parameters with a Markov chain Monte Carlo (MCMC) fitting routine. We carry out MCMC fitting using the Python package \textit{emcee} developed by \citet{Foreman2013}. We use our best-fit orbital elements as a starting point for our 2*N$_{\rm{params}}$ walkers, where the starting point for each walker is perturbed about its best fit value. We assume uniform priors on all of our orbital elements. The quoted error bars on our orbital elements in Table \ref{table:elements} are the standard deviations of the posterior distributions, while the reported orbital element value is the best-fit result from the least-squares routine. 

\section{Detections and New Orbits}
\label{sec:detections}

\subsection{Astrometric Detections with New RVs}
\label{sec:detections1}

After identifying new systems with our ARMADA astrometry data, we followed up with spectra from
the TSU 2m AST to obtain, if possible, RV orbits of the systems. These RV data tell us which component of the outer binary the new companion is orbiting, and it can also confirm the newly detected orbital periods. When we measure the RV motion along with a ``wobble'' motion, we are able to compare the physical and angular $a_1$ term for the inner pair ($a = a_1 + a_2$), where $a_2$ represents the motion of the secondary about the center-of-mass. This gives us an independent measure of the parallax of the system. With a distance and astrometric orbit, we are able to measure the total mass sums of such systems. However, our method does not measure the semi-major axis of the inner orbit in these cases -- only the orbit taken by the brighter primary star (if the flux from the secondary is negligible). We searched for flux from the secondary in all the objects presented in this section, but did not detect the companion. Note that our data are optimized for wide binaries, but follow-up well-calibrated observations could be undertaken to detect or place limits on the flux of inner companions. This means we need an extra piece of information to measure all three masses. We estimate the mass of either the primary or secondary (described in individual subsections) for the stars in this group, which allows us to deduce the other masses in the system. Because we do not have RV information on the outer binary system, there are two possible values for the mutual inclination between the orbits (the equation for mutual inclination is shown in \citealt{muterspaugh2006b}).

Note that we also check both Hipparcos \citep{vanLeeuwen2007} and Gaia \citep{gaia2016} for distances for all objects in this paper. However, we note that as of EDR3, Gaia does not take into account motion of companions when computing parallax. In fact, for all of our ARMADA targets presented here that are listed in Gaia EDR3 the RUWE parameter is $>$1.4, indicating potential bias in the parallax from companions \citep{lindegren2021}. The targets in this paper are also missing from the non-single star catalog of DR3, since they all have binary companions near the edge of Gaia's inner field of view \citep{halbwachs2022}. This means that the Gaia distances for our binaries cannot always be trusted at this time.

\begin{figure}[H]
\centering
\includegraphics[width=7in]{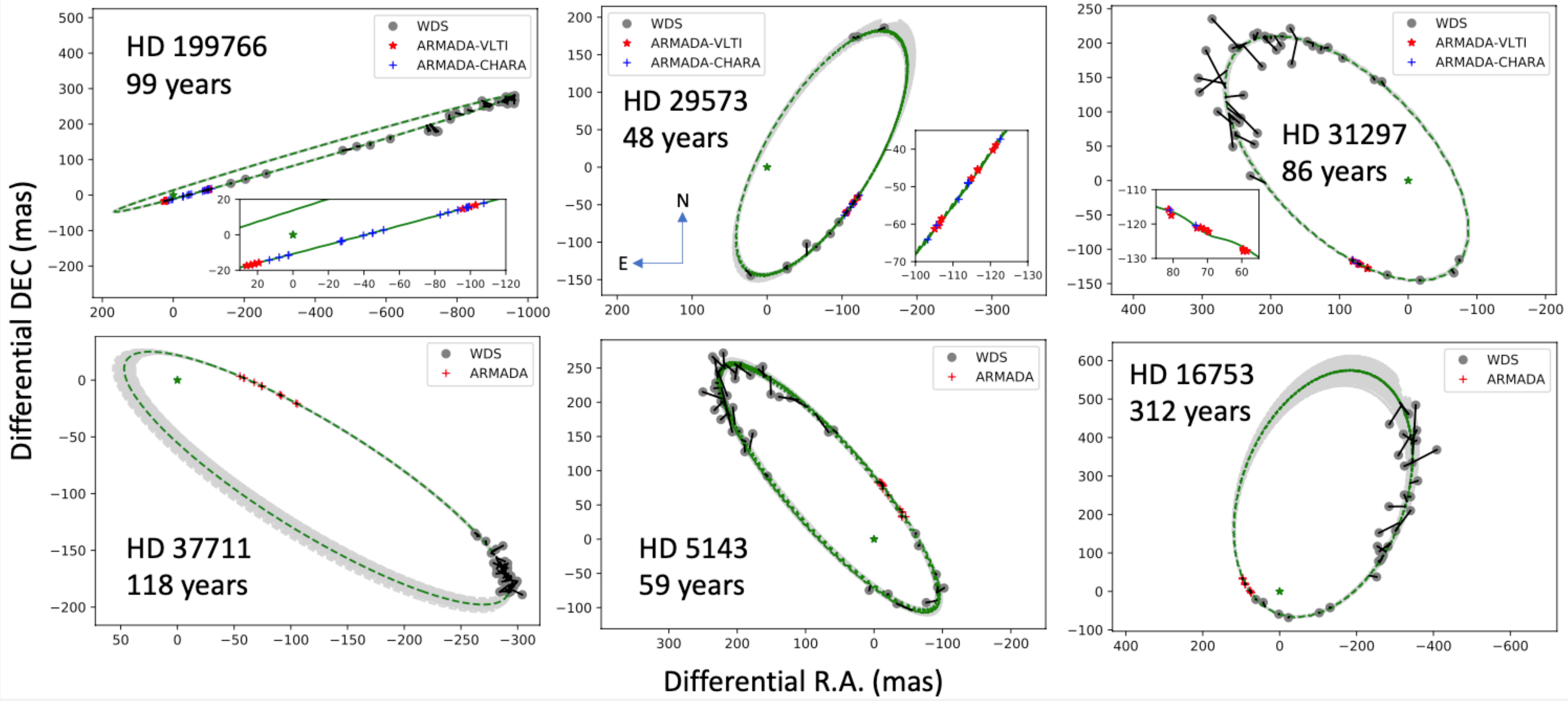}
\caption{We show the outer binary orbits for systems where we detect new companions in astrometric ``wobble" and RV. We combine our new high precision ARMADA epochs with historical speckle data published in WDS. We show the final fitted outer orbit, taking into account the ``wobbles" from newly detected companions in these systems (which can be seen visually at this scale for HD 5143). The shaded grey lines depict 100 random orbits drawn from the posterior of the MCMC chains, while the green dashed line shows the best-fit solution. For the case of HD 37711 and HD 16753, additional orbit monitoring is needed to better constrain the long period orbits. }
\label{outer_orbits1}
\end{figure}

\begin{figure}[H]
\centering
\includegraphics[width=7in]{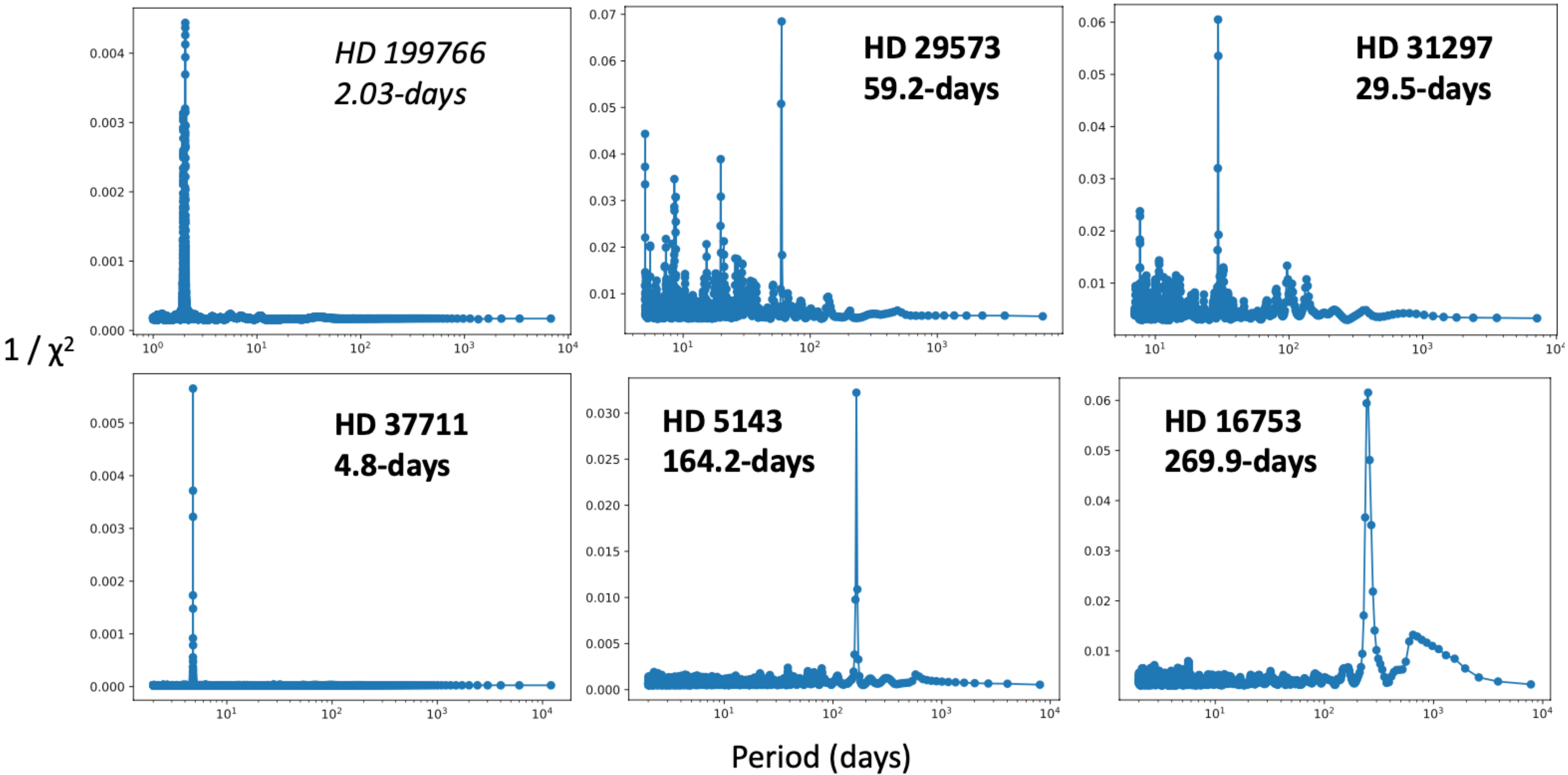}
\caption{We search for additional companions to the wide binary systems of Figure \ref{outer_orbits1} on a grid over inner orbital period. We detect obvious inner periods for four of these six targets. Due to their short periods and small wobbles, we also include RV data for the period searches of HD 199766 and 37711 (the rest are purely astrometric searches). We have RV data on all of these targets to confirm the inner detection. Five of these six targets are new detections (bolded), while HD 199766 is the first astrometric detection of the inner 2-day orbit. }
\label{period_searches1}
\end{figure}

\begin{figure}[H]
\centering
\includegraphics[width=7in]{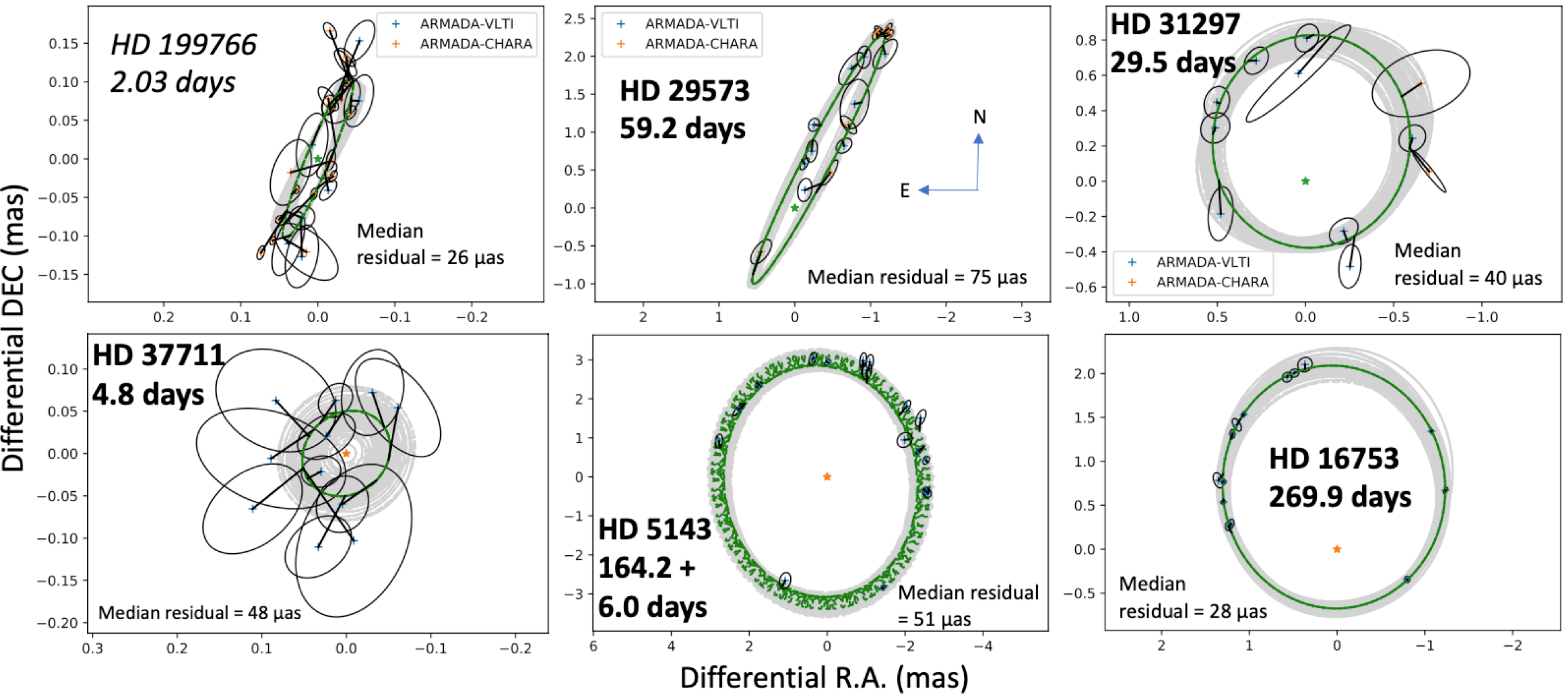}
\caption{We plot the best-fit inner ``wobble" motions due to newly detected companions, with the outer binary motion of Figure \ref{outer_orbits1} subtracted. This motion represents the center-of-light (i.e. the primary star in cases with negligible secondary flux) of the inner orbit orbiting the center-of-mass. The ``wobbles" here range from $\sim$50 $\mu$as to a few mas, with median residuals to the joint RV+astrometry fit of a few tens of $\mu$as. We discover two inner components in HD 5143 (Aa,Ab + Ba,Bb), so in this case the inner orbit measuring Aa to Ba has a ``wobble" motion of its own. The shaded grey regions again depict 100 randomly sampled orbits from the MCMC chains. Since the ``wobble" motion also depends on the subtraction of the outer orbit, the errors on the inner orbits appear larger than expected from the plotted data in some cases. }
\label{inner_orbits1}
\end{figure}

\begin{figure}[H]
\centering
\includegraphics[width=7in]{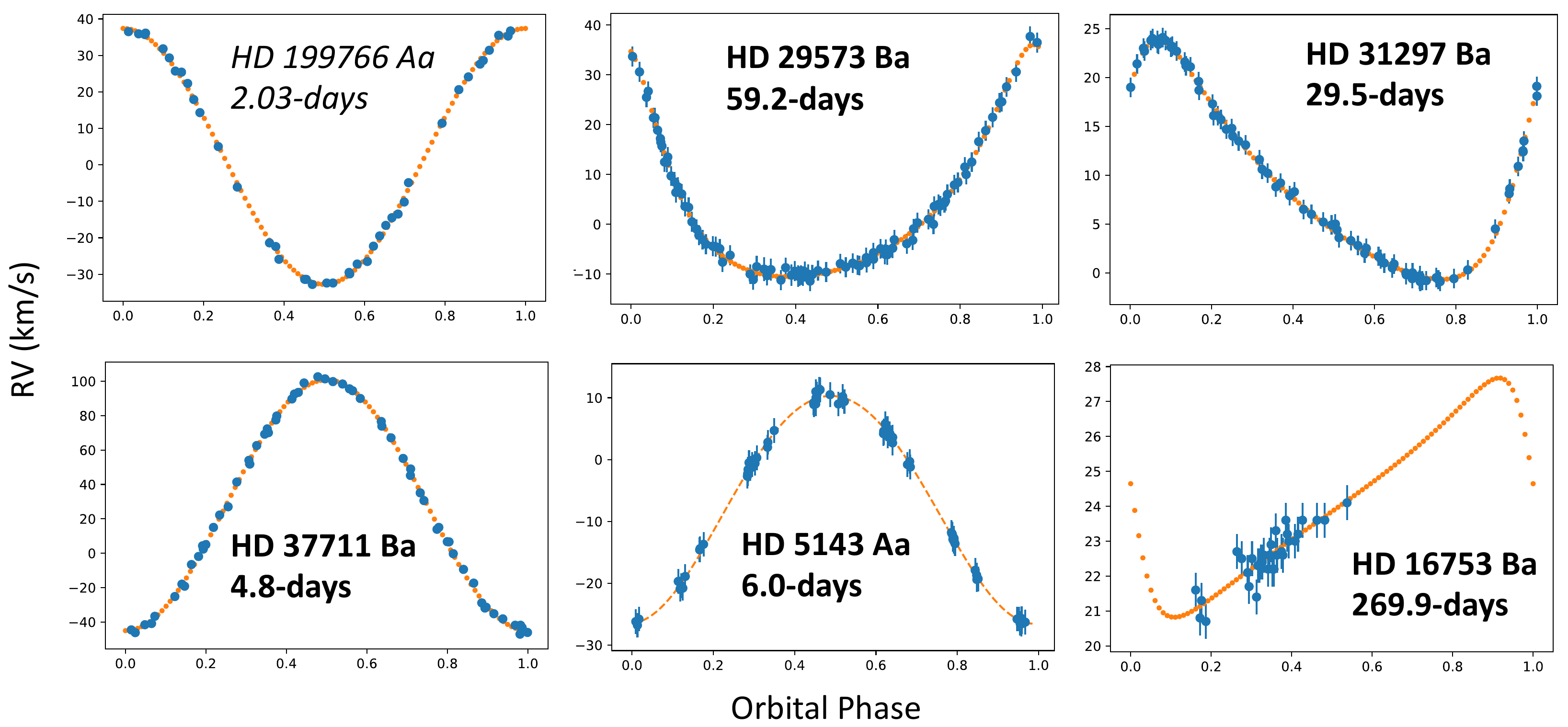}
\caption{We plot the best-fit joint RV inner orbits for companions in Figure \ref{inner_orbits1}, obtained with new TSU-AST spectroscopic data. For each system, we designate the component we are seeing in the RV data (described in individual system notes). These RV data help confirm the newly detected inner orbital elements, tells us which component the companion is orbiting, and allows us to compute a distance. HD 16753 needs further orbit monitoring, due to its long orbital period. }
\label{rv_orbits1}
\end{figure}

We show the final fits to the outer binary orbits for the six objects in this category in Figure \ref{outer_orbits1}, along with 100 randomly sampled orbits from the MCMC chains. Our period grid searches in Figure \ref{period_searches1} reveal the candidate orbital periods for inner companions to these wide binaries with high astrometric residuals. Five of the six targets in this category are new detections, while the detection for HD 199766 is the first astrometric detection of this inner orbit. Finally, we plot the best fit ``wobble" motions for these companions in Figure \ref{inner_orbits1}, along with 100 randomly sampled orbits from the MCMC chains to visually depict errors on the inner orbit. Since the plotting of inner ``wobble" depends on subtraction of the outer orbit that is varying in tandem, these MCMC orbits often appear to fall outside of the plotted error bars. We show the RV fits in Figure \ref{rv_orbits1}. The joint RV+astrometry fits for these systems lead to median residuals to the best-fit orbit of a few tens of micro-arcseconds for both GRAVITY and MIRC-X data. We give notes on individual systems in the following subsections.

\subsubsection{HD 199766}

HD 199766 (eps Equ, HIP 103569) is a known multiple system
with two F-type components, designated A and B, on a 101.5
yr orbit, having high eccentricity and high inclination, 
and angular semi-major axis of 0.647 arcseconds. There 
is a known outer tertiary companion (component C) at 
10.32 arcseconds \citep{tokovinin2017}, and also a 
known inner short-period RV companion, splitting the 
primary into components Aa and Ab. \citet{Abt1976} 
used single-lined RV data to determine an inner Aa,Ab 
orbit with a period of 2.03 days and a semi-amplitude 
of 15.8 km~s$^{-1}$, though the data are somewhat 
sparse and noisy. We obtained 60 new RV measurements 
with the TSU 2~m AST to update this spectroscopic orbit.
Component A is the broader-lined star of the pair
($v \sin i =$ 64 km~s$^{-1}$), and we confirm it to be the 2-day
spectroscopic binary. Component B is also somewhat broad 
lined with $v \sin i =$ 38 km~s$^{-1}$, and during the time 
of our observations it is not significantly variable. 

We also followed this object with both MIRC-X and GRAVITY for the ARMADA survey. Since the interferometric field-of-view for each instrument is near 200 mas, we are not concerned with the wide companion at 10.32 arcseconds. We use the binary motion of the 101.5 yr A,B orbit to measure for the first time the $\sim$100 $\mu$as ``wobble'' motion for the inner spectroscopic orbit. Since our ARMADA data were obtained near periastron passage for this binary system, our precision data help to constrain the outer highly eccentric binary orbit. Table \ref{table:elements} reports our updated orbital elements for HD~199766, along the other systems analyzed in the paper. Our orbital fit fully models the outer binary motion along with the ``wobble'' motion of the inner 2.03 day pair, an orbital period with which we find good agreement with historical RV data from \citet{Abt1976}. However, our semi-amplitude is more than twice as large
as that found by \citet{Abt1976}, likely because they
measured the blended profile.

\begin{figure}[H]
\centering
\includegraphics[width=7in]{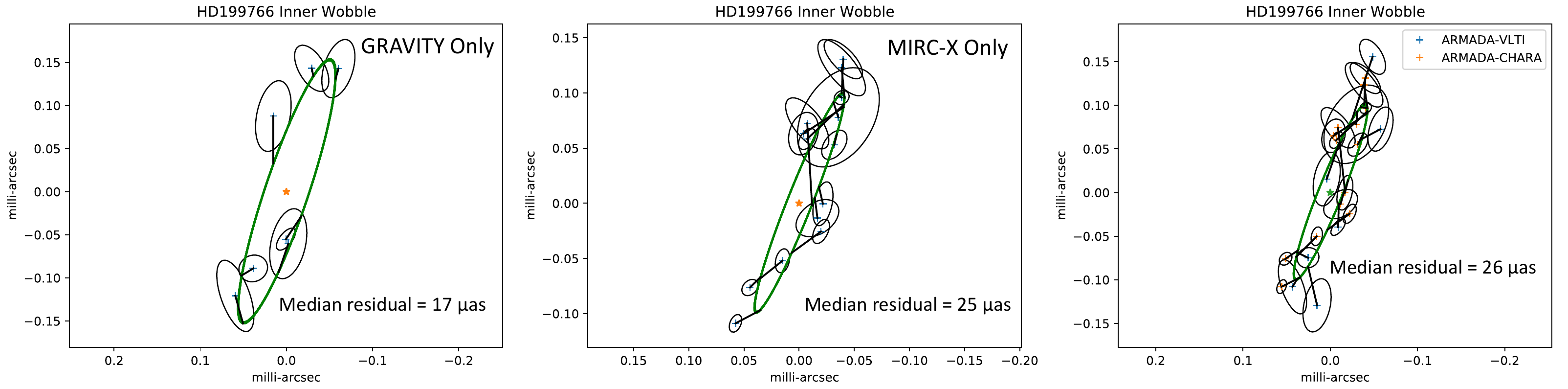}
\caption{Unlike GRAVITY which calibrates to an absolute wavelength scale, our MIRC-X data relies on an internally consistent wavelength calibration scheme. Therefore we need to fit for an extra scale factor when combining MIRC-X+GRAVITY datasets. We show the ``wobble'' fit for GRAVITY data alone (left), MIRC-X data alone (center), and the combined data with a fitted scale factor of $1.00495 \pm 0.00017$ between the instruments (right). In all cases, the inner orbit does not change significantly. }
\label{hd199766_wobbles}
\end{figure}

As mentioned in Section \ref{sec:observations}, the wavelength calibration scheme differs between GRAVITY and MIRC-X data. For MIRC-X, we use our etalon calibration method to bring each night to the same astrometric scale. However, this does not bring our nights to an absolute scale which means there could be up to $10^{-3}$ relative errors in astrometry when combining with other datasets \citep{monnier2012}. On the other hand, GRAVITY uses a wavelength calibration source each night to bring its values to an absolute scale which has an estimated precision at the $5 \times 10^{-5}$ level \citep{sanchez2017}. Since we have both MIRC-X and GRAVITY data on HD 199766, we are able to fit a scale factor between the two sets to bring each ARMADA night on MIRC-X to an absolute wavelength scale. In fact, the inclusion of such a scale factor is needed in order to fit to the combined MIRC-X+GRAVITY data without a systematic offset between the sets. In our triple fit, we simply add a free parameter ``scale factor" which we divide the separations of the MIRC-X datasets by. Figure \ref{hd199766_wobbles} shows the inner astrometric ``wobble'' of our best-fit triple model using the GRAVITY dataset alone, the MIRC-X dataset alone, and the combined datasets. The magnitude and orientation of the ``wobble" is consistent between the two datasets, and we are able to combine them successfully with a best-fit scale factor of $1.00495 \pm 0.00017$ between MIRC-X and GRAVITY (0.5\% shift).

Combining the RV and wobble motions gives a distance of $61\pm4$ pc, which is consistent with the $54\pm4$ pc distance measured by Hipparcos \citep{vanLeeuwen2007}. This object is not listed in Gaia, likely due to its multiplicity. From its absolute Vmag, \citet{tokovinin2018} estimates the B component to be 1.59 M$_{\odot}$. We measure the dynamical mass of the system to be M$_{T} = 4.7\pm0.9$ M$_{\odot}$, which makes M$_{Aa}$ + M$_{Ab}$ = $3.1\pm0.9$ M$_{\odot}$. Hence, by assuming a mass for the B component we can constrain the masses of the inner system to M$_{Aa} = 2.7\pm0.9$ M$_{\odot}$ and M$_{Ab} = 0.45\pm0.09$ M$_{\odot}$. Because we do not have RV information on the outer binary system, there are two possible values for the mutual inclination between the orbits: $52.2\pm2.5^{\circ}$ or $126.8\pm2.5^{\circ}$. Either way, the orbits are not well-aligned between the inner and outer pair.

\begin{small}
\setlength{\tabcolsep}{0.1pt}
\begin{longtable}{lccccccccc}
\caption{Best Fit Orbital Elements}
\label{table:elements}\\
\hline
HD (component) & $P$ & $a$ (mas) \tablenotemark{a} & $e$ & $i$ ($^\circ$) & $\omega$ ($^\circ$) & $\Omega$ ($^\circ$) & $T$ (MJD) & $K$ (km s$^{-1}$) & $\gamma$ (km s$^{-1}$) \\
\hline
\endhead

199766 (A,B) & 98.0 yr & 590.6 & 0.712 & 92.286 & 350.7 & 106.024 & 60608 & -- & -- \\
& $\pm$1.1 & $\pm$2.2 & $\pm$0.003 & $\pm$0.019 & $\pm$0.9 & $\pm$0.023 & $\pm$13 & & \\
\\
199766 (Aa,Ab) & 2.03121 d & 0.112 & 0.0 & 97.7 & 0.0 & 157.2 & 59368.456 & 35.1 & 2.30 \\
& $\pm$0.00017 & $\pm$0.003 & -- & $\pm$1.6 & -- & $\pm$2.4 & $\pm$0.004 & $\pm$0.5 & $\pm$0.02 \\
 
\hline

29573 (A,B) & 47.9 yr & 303 & 0.782 & 74.8 & 282.3 & 154.2 & 53233 & -- & -- \\
& $\pm$0.8 & $\pm$10 & $\pm$0.009 & $\pm$0.6 & $\pm$0.6 & $\pm$0.4 & $\pm$88 & & \\
\\
29573 (Ba,Bb) & 59.18 d & 1.88 & 0.407 & 84 & 20 & 152 & 58774.2 & 23.3 & 3.8 \\
& $\pm$0.03 & $\pm$0.05 & $\pm$0.013 & $\pm$1 & $\pm$2 & $\pm$1 & $\pm$0.4 & $\pm$0.5 & $\pm$0.2 \\

\hline

31297 (A,B) & 85.7 yr & 243.3 & 0.525 & 120.4 & 129.2 & 38.4 & 53119 & -- & -- \\
& $\pm$0.4 & $\pm$1.1 & $\pm$0.002 & $\pm$0.2 & $\pm$0.3 & $\pm$0.3 & $\pm$40 & & \\
\\
31297 (Ba,Bb) & 29.481 d & 0.61 & 0.379 & 165 & 306 & 119 & 59546.94 & 12.28 & 8.84 \\
& $\pm$0.011 & $\pm$0.04 & $\pm$0.013 & $\pm$10 & $\pm$3 & $\pm$3 & $\pm$0.16 & $\pm$0.17 & $\pm$0.13 \\

\hline

37711 (A,B) & 118 yr & 185 & 0.78 & 68.9 & 338 & 63.5 & 61118 & -- & -- \\
& $\pm$7 & $\pm$7 & $\pm$0.01 & $\pm$0.7 & $\pm$4 & $\pm$0.8 & $\pm$35 & & \\
\\
37711 (Ba,Bb) & 4.77095 d & 0.055 & 0.0 & 30 & 0.0 & 310 & 59529.10 & 72.9 & 27.9 \\
& $\pm$0.00023 & $\pm$0.013 & -- & $\pm$19 & -- & $\pm$14& $\pm$0.01 & $\pm$0.3 & $\pm$0.2 \\

\hline

\\

5143 (A,B) & 58.8 yr & 233.6 & 0.429 & 73.4 & 179.8 & 41.16 & 56223 & -- & -- \\
& $\pm$0.3 & $\pm$1.4 & $\pm$0.004 & $\pm$0.2 & $\pm$0.6 & $\pm$0.13 & $\pm$40 & & \\
\\
5143 (Aa,Ab) & 6.0008 d & 0.16 & 0.0 & 153 & 0.0 & 239 & 59522 & 18.39 & -7.5 \\
& $\pm$0.0006 & $\pm$0.03 & -- & $\pm$30 & -- & $\pm$12 & $\pm$10 & $\pm$0.18 & $\pm$0.12 \\
\\
5143 (Ba,Bb) & 164.24 d & 3.13 & 0.07 & 33 & 292 & 8 & 59476 & -- & -- \\
& $\pm$0.13 & $\pm$0.04 & $\pm$0.02 & $\pm$2 & $\pm$16 & $\pm$3 & $\pm$7 &  &  \\

\hline

16753 (A,B) & 314 yr & 389 & 0.833 & 147.9 & 256.8 & 50.4 & 57696 & -- & -- \\
& $\pm$11 & $\pm$8 & $\pm$0.004 & $\pm$0.45 & $\pm$1.9 & $\pm$1.7 & $\pm$12 & & \\
\\
16753 (Ba,Bb) & 271.1 d & 1.48 & 0.51 & 159 & 85 & 267 & 58339 & 3.4 & 24 \\
& $\pm$1.2 & $\pm$0.04 & $\pm$0.03 & $\pm$6 & $\pm$21 & $\pm$21 & $\pm$5 & $\pm$1.4 & $\pm$1 \\

\hline
 
1976 (A,B) & 171 yr & 208.0 & 0.162 & 62.8 & 306.0 & 27.8 & 33710 & -- & -- \\
& $\pm$3 & $\pm$2.7 & $\pm$0.008 & $\pm$0.4 & $\pm$4 & $\pm$0.4 & $\pm$679 & & \\
\\
1976 (Aa,Ab) & 25.4163 d & 0.42 & 0.05 & 100.7 & 61 & 339.9 & 59477 & 24.2 & -7 \\
& $\pm$0.0008 & $\pm$0.05 & $\pm$0.03 & $\pm$1.2 & $\pm$52 & $\pm$1.1 & $\pm$5 & $\pm$2.3 & $\pm$4 \\
 & & 1.08 (Aa,Ab) & & & & & & & \\
& & $\pm$0.02 & & & & & & & \\

\hline

173093 (A,B) & 7.230 yr & 82.14 & 0.6041 & 104.20 & 255.76 & 287.60 & 55471.9 & 10.07 (A) & -49.017\\
 & $\pm$0.003 & $\pm$0.15 & $\pm$0.0009 & $\pm$0.04 & $\pm$0.12 & $\pm$0.04 & $\pm$0.6 & $\pm$0.03 & $\pm$0.018 \\
 &  &  &  &  &  &  &  & 20.312 (B) & \\
  &  &  &  &  &  &  &  & $\pm$0.048 & \\
\\
173093 (Aa,Ab) & 2.3580109 d & 0.313 & 0.0 & 15 & 0.0 & 116.7 & 53853.9425 & 21.64 (Aa) & -49.017 \\
 & $\pm$0.0000014 & $\pm$0.008 & -- & $\pm$5 & -- & $\pm$0.7 & $\pm$0.0014 & $\pm$0.03 & $\pm$0.018 \\
 &  & 0.648 (Aa,Ab) &  &  &  &  &  & 22.83 (Ab) &  \\
 &  & $\pm$0.006 &  &  &  &  &  & $\pm$0.05 &  \\
 
 \hline
 
220278 (A,B) & 64.30 yr & 407 & 0.142 & 77.53 & 178.3 & 96.01 & 53870 & -- & -- \\
-- & $\pm$0.27 & $\pm$2 & $\pm$0.003 & $\pm$0.10 & $\pm$1.6 & $\pm$0.06 & $\pm$107 & & \\
\\

220278 (Aa,Ab) & 962.2 d & 11.18 & 0.0518 & 79.16 & 189 & 80.61 & 59269 & -- & -- \\
-- & $\pm$1.4 & $\pm$0.03 & $\pm$0.0019 & $\pm$0.13 & $\pm$4 & $\pm$0.15 & $\pm$11 & & \\
& & 41.90 (Aa,Ab) & & & & & & & \\
& & $\pm$0.13  & & & & & & & \\

\hline

196088/9 (A,B) & 116 yr & 102.1 & 0.492 & 51.6 & 56 & 52.3 & 45701 & -- & -- \\
& $\pm$4 & $\pm$2.9 & $\pm$0.016 & $\pm$1.2 & $\pm$4 & $\pm$1.8 & $\pm$135 & & \\
\\
196088/9 (Aa,Ab) & 46.7 d & 0.072 & 0.0 & 64 & 0.0 & 142 & 59140 & -- & -- \\
& $\pm$0.2 & $\pm$0.013 & -- & $\pm$21 & -- & $\pm$47 & $\pm$6 & & \\

\hline

48581 (A,B) & 84 yr & 162 & 0.7 & 97 & 358 & 179 & 63320 & -- & -- \\
& $\pm$7 & $\pm$15 & $\pm$0.2 & $\pm$2 & $\pm$3 & $\pm$4 & $\pm$54 \\
\\
\\
48581 (Ba,Bb) & 4.8385 d & 0.214 & 0.0 & 62 & 0.0 & 122 & 58383.0 & -- & -- \\
& $\pm$0.0028 & $\pm$0.036 & -- & $\pm$17 & -- & $\pm$22 & $\pm$0.3 & & \\

\textit{48581 (Ba,Bb)}\tablenotemark{b} & \textit{10.415 d} & \textit{0.27} & \textit{0.0} & \textit{75} & \textit{0.0} & \textit{114} & \textit{59259.9} & -- & -- \\
& \textit{$\pm$0.006} & \textit{$\pm$0.04} & & \textit{$\pm$20} & & \textit{$\pm$25} & \textit{$\pm$0.4} & & \\

\hline

185762 (A,B) & 20.53 yr & 89.9 & 0.072 & 136.1 & 3 & 18.9 & 57922 & -- & -- \\
& $\pm$0.17 & $\pm$1.8 & $\pm$0.010 & $\pm$1.3 & $\pm$9 & $\pm$1.9 & $\pm$178 & & \\
\\
185762 & 6.9 yr & 10 & 0.529 & 111.4 & 298 & 143.3 & 59528 & -- & -- \\
(Aa,Ab? Ba,Bb?) & $\pm$0.2 & $\pm$0.8 & $\pm$0.021 & $\pm$2.1 & $\pm$5 & $\pm$1.3 & $\pm$20 & & \\

\hline

\end{longtable}
\tablenotetext{a}{Denotes the semi-major axis of the \textit{wobble} motion for the inner orbits, unless otherwise noted. It is the binary semi-major axis for outer orbits.}
\tablenotetext{b}{Second best solution, from multiple peaks in periodogram}
\end{small}

\subsubsection{HD 29573}
HD 29573 (HIP 21644) is a bright A-type star that was discovered to be a binary by the Hipparcos mission \citep{vanLeeuwen2007}. \citet{cvetkovi2014} characterized its binary orbit further with newly obtained speckle data. Despite the low number of data points on the 40 year period, they were able to compute a dynamical mass of 4.99$\pm$0.73 M$_{\odot}$, which was somewhat consistent within error bars with the photometric mass obtained from the spectral types of the two stars in the system -- spectral types A1 and F2 (corresponding to a photometric system mass of 3.84 M$_{\odot}$). Though error bars are high, it did seem that there was some extra mass in the system from the higher dynamical mass compared to the photometric mass. We obtained new epochs on this system with the ARMADA survey, both at CHARA and and VLTI. Although our data did not extensively improve the orbital coverage of the outer system, our high precision astrometric data allowed us to search for extra companions in the binary system. 

We detect a clear ``wobble'' motion with an inner period of $\sim$60 days, with a consistent period found searching the CHARA and the VLTI datasets separately. The $\sim$1.4 mas wobble motion makes this a mass which is clearly stellar, and so we followed up the system with new RV data obtained with the TSU 2m AST. We see two components with both A-star and solar line lists. The lines of the primary are stronger and broader, having a $v$~sin~$i$ of 28 km~s$^{-1}$ and this
component has an apparently constant RV. The secondary component of the binary is weaker but narrow-lined with a $v$~sin~$i$ of 8.4 km~s$^{-1}$ and is clearly RV variable. Hence, our new ``wobble'' orbit is describing the orbit of the B component of the binary system -- making the configuration A+Ba,Bb. Our RV data independently confirm the newly detected inner period, and hence we are able to 
perform a full joint fit on this system. 

\begin{figure}[h]
\centering
\includegraphics[width=7in]{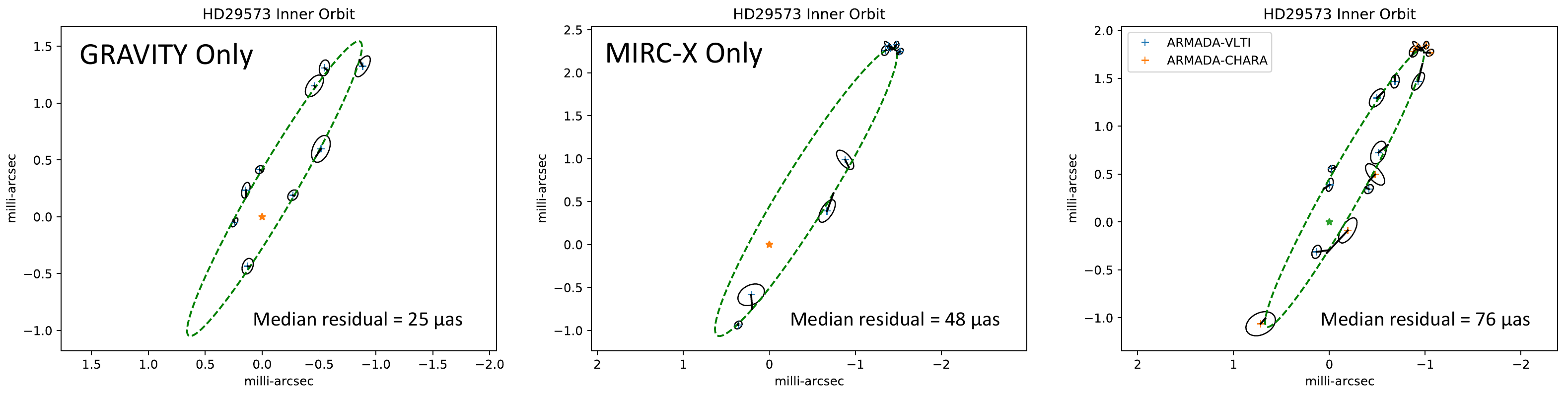}
\caption{Same as Figure \ref{hd199766_wobbles}, but for HD 29573. We show the fit to the inner companion with GRAVITY alone (left), MIRC-X alone (center), and the combined fit with a fitted scale factor of 1.00505 (right). This value agrees well with that obtained on HD 199766. }
\label{hd29573_wobble}
\end{figure}

As for the case of HD 199766, we have both MIRC-X and GRAVITY data for this object. Hence we are able to fit for a scale factor between the two sets, which is needed to combine the sets but also allows us to fit for an astrometric scale factor between the absolute wavelength scale of GRAVITY and the relative scale achieved with the etalons on MIRC-X. Figure \ref{hd29573_wobble} shows the inner wobble fit for GRAVITY data alone, MIRC-X data alone, and the combined sets. The best-fit scale factor between the two sets is 1.00501 $\pm$ 0.00021 (0.5\% shift), which is consistent with our fitted value for HD199766. However, in Figure \ref{hd29573_wobble} it is apparent that the MIRC-X-only orbit prefers a slightly higher eccentricity than the GRAVITY dataset. Since orbital coverage is different between the two sets, this may have an effect on the astrometry-only fits. When combining with our RV1 orbit in Figure \ref{inner_orbits1} we are confident that we report the correct eccentricity, although residuals are higher on this system than for the shared source HD 199766. 

We measure a distance of 62.0 $\pm$ 2.1 pc, which is lower than the Gaia DR3 measurement of 70.3 $\pm$ 1.8 pc \citep{gaia_dr3}. We note that Gaia DR2 (61.57 $\pm$ 1.06 pc, \citealt{gaia_dr2}) is consistent with our measurement, though Hipparcos (71.3 $\pm$ 2.8 pc, \citealt{vanLeeuwen2007}) is also higher. Our lower distance leads to a total system mass sum of 2.4 $\pm$ 0.4 M$_{\odot}$, which is too low for the spectral types. This likely indicates that the outer orbit is not yet well constrained. Alternatively, there could be a mismatch between the components we are viewing in the RV and astrometry (Ba/Bb) which would lead to an incorrect distance. We instead use the distance computed in Gaia DR3 and consistent with Hipparcos. This leads to a total mass sum of the system of 4.2 $\pm$ 0.5 M$_{\odot}$. \citet{kervella2022} estimated a mass of 2.65 M$_{\odot}$ for the primary, from the isochrones by \citet{girardi2000} and following the procedure of \citet{kervella2019}. Assuming a mass of 2.65 M$_{\odot}$ for the primary, we find M$_{Ba}$ + M$_{Bb}$ = 1.6 $\pm$ 0.5 M$_{\odot}$. Individual masses from the ``wobble" motion are 1.0 $\pm$ 0.4 M$_{\odot}$ and 0.60 $\pm$ 0.15 M$_{\odot}$ for Ba and Bb, respectively. The inner and outer orbits have similar inclination values, and the mutual inclination between the orbits is either 160.2$^{\circ}$$\pm$0.7$^{\circ}$ or 9.0$^{\circ}$$\pm$0.7$^{\circ}$.

\subsubsection{HD 31297}

HD 31297 (HIP 22812) is a known early F-type visual binary system. A first attempt at a full orbit characterization was made by \citet{seymour2002}, using speckle data for the visual binary orbit. Their period estimate of 147 years was more recently updated by \citet{malkov2012}, using all data compiled from the WDS and ORB6 catalogs. The spectral type of the primary was reported as F5, with an orbital period of 84 years for the binary. They computed a dynamical mass of 3.79$\pm$1.45 M$_{\odot}$, which was consistent with the photometric mass estimate of 3.77 M$_{\odot}$. We followed this system with VLTI-GRAVITY, which quickly showed large residuals to the pure binary fit. We detect a clear inner orbital period of $\sim$30 days, with a ``wobble'' semi-major of about 0.7 mas. This large wobble implies a stellar mass companion, and indeed we can detect this inner period with new RV data from the TSU 2m AST. 

Two components are seen in the spectrum with both the A-star and solar line list, 
though both are better defined with the solar line list and have somewhat similar equivalent
widths. The primary component has the broader lines with a $v$~sin~$i$ of 80 km~s$^{-1}$, and 
we confirm this star to have constant RV. Hence it is the secondary component with a $v$~sin~$i$
of 20 km~s$^{-1}$ that is ``wobbling'' -- making the system configuration A+Ba,Bb. We 
independently confirm the newly detected period with our RV curve. 

We also obtained two MIRC-X epochs on this target, which gives us another opportunity to compute the wavelength scale factor between the two instruments (as for HD 199766 and HD 29573). Unfortunately, both MIRC-X epochs were taken under poor conditions and have rather large error ellipses as can be seen in Figure \ref{inner_orbits1}. Still, they do fit the outer and inner orbits well with a scale factor between MIRC-X and GRAVITY of $1.0061\pm0.0005$ (0.6\%). This is slightly larger than the correction computed from the other two objects, though the higher error bar here makes it consistent within $<$2 sigma. 

Since the ``wobble" motion of the newly detected companion is small ($<$1 mas), our fractional error on the physical motion of the Ba component is large (computed with RV semi-amplitude, inner orbital period, inclination, and eccentricity). This makes our distance error very large at 200 $\pm$ 130 pc. Hence, to compute masses we assume the Hipparcos distance of 125 $\pm$ 11 pc \citep{vanLeeuwen2007}. Gaia does not list a parallax measurement for this system, likely due to its binarity. Using this distance and our orbital elements, we compute a total mass of 3.8 $\pm$ 1.0 M$_{\odot}$ for the system. From its spectral type and effective temperature, \citet{reiners2020} report a mass of 1.4 M$_{\odot}$ for the primary. Assuming this mass for the primary, we compute M$_{Ba}$ + M$_{Bb}$ = 2.4 $\pm$ 1.0 M$_{\odot}$. Our ``wobble" orbit then implies masses of 1.7 $\pm$ 0.7 M$_{\odot}$ and 0.74 $\pm$ 0.27 M$_{\odot}$ for Ba and Bb components respectively. Since the A component is brighter, it is likely that the true mass of the Ba component is on the low end of its 1-sigma error bar. The inner and outer orbits are not very well aligned, with a mutual inclination of either 58.6 $\pm$ 0.4$^{\circ}$ or 63.3 $\pm$ 2.8$^{\circ}$.

\subsubsection{HD 37711}
HD 37711 (126 Tau, HIP 26777) is a B-type star with a known visual companion. \citet{docobo1999} found a period of 115 
yr for the highly eccentric orbit. Using speckle data compiled from WDS, \citet{malkov2012} confirmed this orbital period for the B8V+B7V system, with a high eccentricity of 0.87. However, only a small arc of this orbit far from periastron has been covered, which makes these orbital elements more uncertain because of
the high eccentricity. The dynamical mass of the binary was reported as 17.71 M$_{\odot}$, with an extremely high uncertainty of about 24 M$_{\odot}$. This uncertainty makes it impossible to compare to the reported photometric mass estimate of 9.78 M$_{\odot}$, though the authors also report a lower 6.41 M$_{\odot}$ spectroscopic mass estimate. The mass inconsistencies likely result from either poor orbital coverage, poor photometry or spectral type estimates, or extra companions in the system. We followed this system with ARMADA using CHARA-MIRC-X data, and our observations fall closer to periastron passage, coverage of which is important for improving the outer binary orbit. In addition, we detect high residuals to the binary fit, which could imply extra short period companions in the system. With ARMADA data alone, we detect a few short period orbits that fit the ``wobble'' motion (Figure \ref{period_searches1}). To confirm the inner period, we followed this system with new RV data.

The primary component of HD 37711 has rather broad lines with a $v$~sin~$i$ of 83 km~s$^{-1}$ that we have measured with a B-star line list. While the RVs have a small range of velocity variability, about 8 km~s$^{-1}$, because of the very broad lines, we believe that this component
is nearly constant in RV. With our blueward A star line
list, we can detect a second component, which has very weak ($<$1\% deep) and narrow spectral lines ($v$~sin~$i$ = 6 km~s$^{-1}$). From our 62 measured RVs, we confirm an inner period for the system. This velocity variation resulted in a short orbital period of 4.77 days and a circular orbit. Though we attribute this RV variation to the Ba component, it is also possible we are measuring the secondary of the spectroscopic binary here (the Bb component). We again are able to constrain the configuration of the newly detected system to A+Ba,Bb. 

Because its $\sim$50 $\mu$as wobble and high error bars due to the low flux ratio of the outer binary, our data are not precise enough to measure the distance independently. Gaia DR3 lists a distance of 720 $\pm$ 310 pc for this system \citep{gaia_dr3}, which is likely very uncertain due to the multiplicity of the system. Hipparcos did a bit better, with a distance of 195 $\pm$ 31 pc \citep{vanLeeuwen2007}. This distance also leads to a more consistent dynamical mass for the system of 4.6 $\pm$ 2.3 M$_{\odot}$, although the error bars on this value make it impossible to constrain individual masses without a more precise distance, a better constrained outer orbit, and lower errors for the inner orbit. Since the system consists of two B-type stars, the true mass sum is likely on the high end of our range. Though our inner inclination error is high, we are able to constrain the mutual inclination between the outer and inner orbits to 60$^{\circ}$$\pm$14$^{\circ}$ or 82$^{\circ}$$\pm$32$^{\circ}$, implying that the two orbits are not well aligned.

\subsubsection{HD 5143}
HD 5143 (HIP 4176) is a known early F-type visual binary system. Using data from WDS, \citet{malkov2012} found an orbital period of 57.67 yr for the orbit, with an eccentricity of 0.52. The estimated dynamical mass from that work is 7.12$\pm$2.18 M$_{\odot}$, which is quite far from the photometric mass estimate of 3.24 M$_{\odot}$. This implies either an incorrect orbit (although orbital coverage from WDS is quite good), poor photometry, or potentially additional components in the system. We followed this binary with VLTI-GRAVITY over a part of the outer orbit that was not covered by speckle. Along with improving the confidence of the outer orbital elements, our precision data unveiled additional ``wobble'' motion in the system due to previously unseen companions. We see a clear period peak at 164 days, with a wobble semi-major axis of 3.1 mas. This large wobble implies a new stellar mass component in the system. However, our best triple orbit still displays high residuals which hints at a potential fourth object in the system. With 21 free parameters for a 4-body Keplerian orbit, it is not easy to identify the additional peak in period space with a low number of data points from astrometry. 

To aid in our search, we followed up on this object with new RV data. We see two components in the spectrum, both with variable RVs. The lines of the primary star are narrower, $v$~sin~$i$ = 10 km~s$^{-1}$, and much deeper than those of the secondary , $v$~sin~$i$ = 29 km~s$^{-1}$, of the visual binary, and the lines of both components are significantly blended with each other. The RVs of the primary are easily measured, and display a 6.0-day period with a circular orbit. This is not the dominant period seen in the astrometry, so we also searched for an RV signal around the fainter secondary of the outer binary system. We indeed are able to see a long period variation around this star, confirming the configuration of the system to Aa,Ab + Ba,Bb. However,
our measurements of the RVs for lines of the much weaker visual
binary secondary appear to be contaminated by the much stronger
lines of the primary. A period grid search shows two possible
periods, one at $\sim$153 days and a less likely period at 6 days, the latter being the same as that found for the primary. Thus, while the RVs of the secondary clearly have a long-term variation, we are not able to obtain a well-fit Keplerian orbit. We conclude that the contamination from the primary is likely corrupting our RV measurements for the weak and broader lined visual secondary. 

Although our RV data tells us the configuration of the system is Aa,Ab (6-day) + Ba,Bb (164-day), with only one RV orbit out of the four components we are not able to disentangle all of the masses. The distance from Hipparcos for the system is 126$\pm$12 pc \citet{vanLeeuwen2007}, which agrees within 2-sigma with the Gaia distance of 103.1$\pm$1.8 pc \citet{gaia_dr3}. Since we measure the small ``wobble" from the 6-day spectroscopic orbit, we compute a distance of 90$\pm$26 pc, which is in better agreement with Gaia although the error bar is large. If we take the Gaia distance we compute a total system mass of 4.05$\pm$0.23 M$_{\odot}$, though we are unable to compute individual masses without additional RV data due to each component being a binary itself.  The 6-day orbit of Aa,Ab has a mutual inclination with the outer binary of either 80$^{\circ}$ or 120$^{\circ}$ $\pm$ 30 $^{\circ}$, while the Ba,Bb orbit has a mutual inclination of the outer orbit of either 101.6$\pm2.0^{\circ}$ or 47.2$\pm2.0^{\circ}$. In either case, the inner orbits are not well aligned with the outer long period binary.

\subsubsection{HD 16753}
HD 16753 (HIP 12466) is a mid F-type binary system with an orbit first determined by \citet{ling2010}, who reported a period of 271.7 yr and high eccentricity of 0.833. They computed a dynamical mass of 4.2$\pm$2.0 M$_{\odot}$ and argue that the spectral type is perhaps better classified as A-type rather than the F5V reported by WDS and SIMBAD. \citet{malkov2012} reported a reduced orbital period for this system of 212.5 yr, though there is no dynamical mass estimate likely due to poor orbital coverage. Our ARMADA data from VLTI-GRAVITY improve the outer elements slightly, being located near periastron passage of the high eccentricity orbit. However, there is still a large fraction of the orbit that is uncovered by speckle data, and there are no observations near apastron. Our high residuals to the binary fit imply an additional ``wobble'' motion due to a newly discovered companion, and we see a strong period peak at 251 days and 1.4 mas for the semi-major axis wobble motion. The companion is of stellar mass, so we attempted to follow up these observations with new RV. 

HD 16753 shows two sets of lines in our newly obtained spectra. They have similar broadenings and sit almost directly on top of each other, making the individual RVs somewhat difficult to measure. The stronger lined star, presumably the primary, has a $v$~sin~$i$ of 39 km~s$^{-1}$ while the secondary has a $v$~sin~$i$ of 24 km~s$^{-1}$. The primary component is constant in RV over our of 100 days of 
observation. The secondary is clearly varying in RV with a constant slope. As can be seen in Figure \ref{rv_orbits1}, more spectra are needed to define the RV orbit and decrease the uncertainty in velocity semi-amplitude. Though we are able to confirm that the 251-day period is consistent with the RV variation shown so far, and the B-component of the wide binary hosts the newly detected companion (making the configuration A + Ba,Bb). 

Since our RV curve is incomplete for this system, our distance has a high error of 140$\pm$40 pc. This is within 2-sigma from the Gaia DR3 value of 198.3 $\pm$ 3.1 pc \citep{gaia_dr3}. We also note that Hipparcos has high error bars as well, which makes it consistent with our measurement at 188 $\pm$ 29 pc. Taking the precision Gaia measurement to be correct, we compute a total mass sum of 4.7 $\pm$ 0.5 M$_{\odot}$. \citet{kervella2019} report a mass of 2.063 M$_{\odot}$ for the primary using isochrones from \citet{girardi2000}. This mass estimate also supports the argument from \citet{ling2010} that the primary is better classified as an A-type star. Assuming the mass for the primary, we compute a mass sum M$_{Ba}$ + M$_{Bb}$ = 2.6$\pm$0.5 M$_{\odot}$. Individual masses for Ba and Bb are then 1.9$\pm$0.4 M$_{\odot}$ and 0.68$\pm$0.09 M$_{\odot}$, respectively. The mutual inclination of this system is either 19$^{\circ}$$\pm$4$^{\circ}$ or 50$^{\circ}$$\pm$4$^{\circ}$. 

\subsection{Systems with Inner Visual Component Detected}
\label{sec:detections2}

\begin{figure}[H]
\centering
\includegraphics[width=7in]{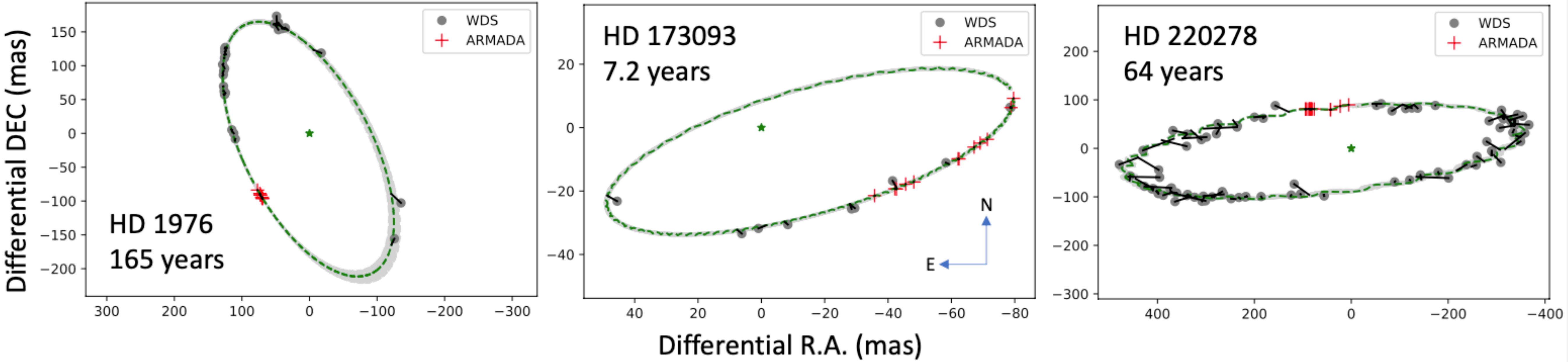}
\caption{We show the outer binary orbits for systems where we directly detect the flux from inner visual components in our interferometric data. We combine our new high precision ARMADA epochs with historical speckle data published in WDS. We show the final fitted outer orbit, taking into account the ``wobbles" from newly detected companions in these systems. In the case of HD 220278, the ``wobble'' from its newly detected companion is at the 10 mas level, which makes it extremely obvious even at the wide scale. }
\label{outer_orbits2}
\end{figure}

\begin{figure}[H]
\centering
\includegraphics[width=7in]{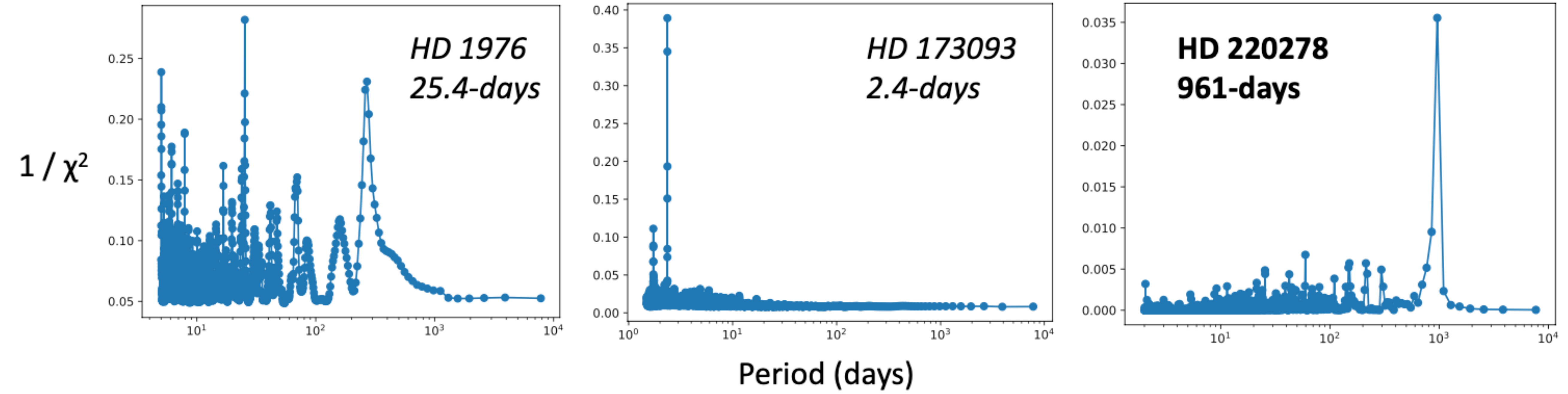}
\caption{After our initial guess at the best fit outer orbit, we search for additional astrometric companions on a grid over inner orbital period for the systems in Figure \ref{outer_orbits2}. These systems all show rather obvious inner orbital periods, with HD 1976 being confirmed with RV data to be on the correct peak. HD 220278 is a new detection, with a long period and large ``wobble" amplitude. }
\label{period_searches2}
\end{figure}

\begin{figure}[H]
\centering
\includegraphics[width=7in]{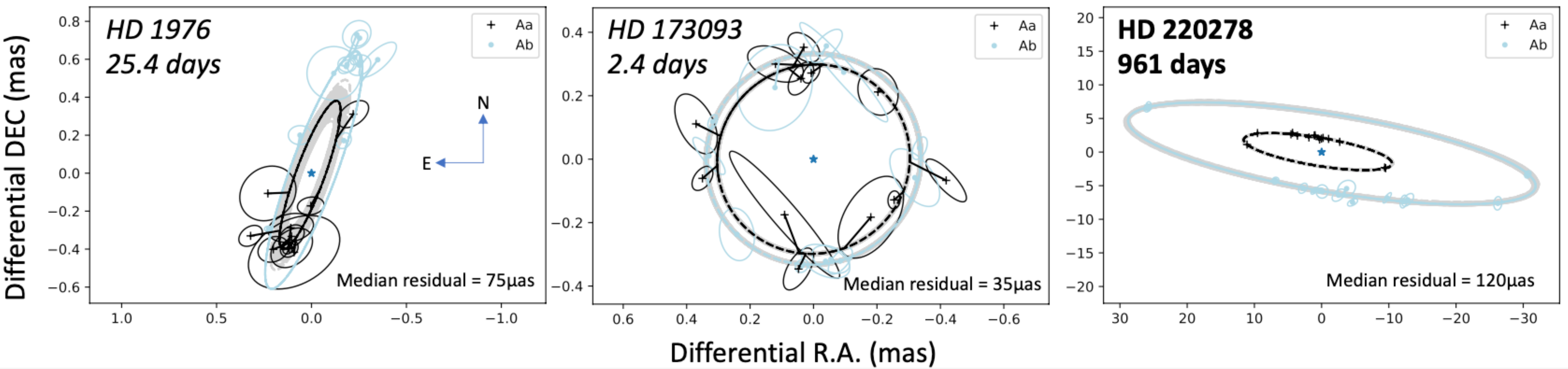}
\caption{We plot the best-fit inner orbits due to the newly detected companions for the systems in Figure \ref{outer_orbits2}. In these cases we see the flux from both inner components, so we plot the motion of each one around its center-of-mass. The median residuals to the orbit fits are a bit higher for HD 1976 and HD 220278, which are likely due to higher measurement uncertainties on low flux inner companions. In the case of HD 220278, there is likely a fourth component in the system which has a yet unconstrained orbit. The grey shaded regions depict 100 random orbits from the MCMC chains, though for HD 1976 we only show the MCMC orbits for the inner component for clarity since the two orbits overlap.}
\label{inner_orbits2}
\end{figure}

\begin{figure}[H]
\centering
\includegraphics[width=7in]{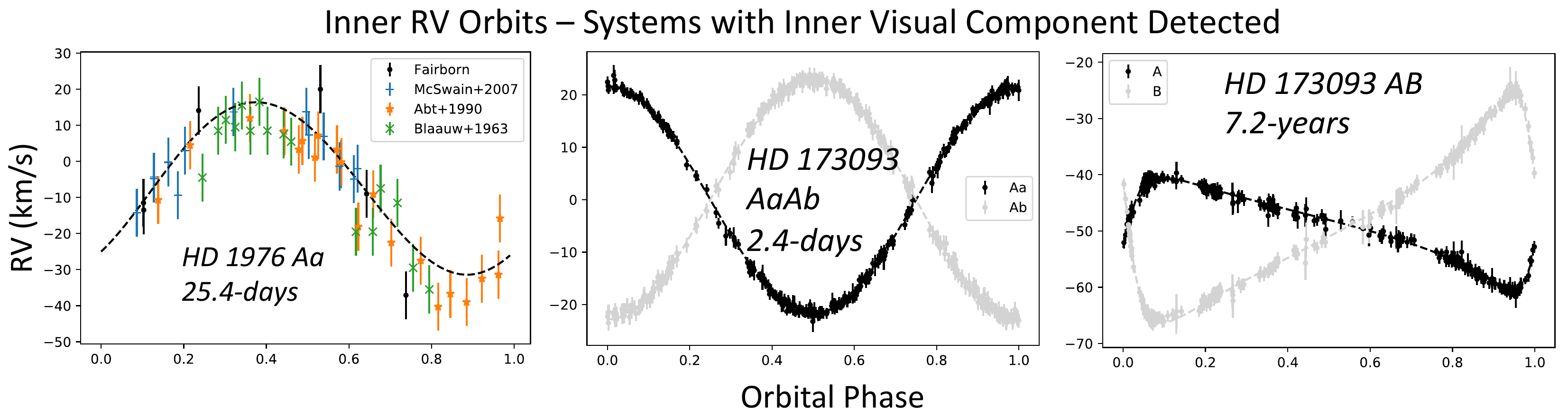}
\caption{(Left) We combined published RV data from \citet{mcswain2007,abt1990,blaauw1963} with 5 new epochs taken at Fairborn Observatory. The single-line RV curve is noisy here, due to the wide and blended profiles of this B-type system. (Center and Right) We combined our new astrometry of HD173093 with the triple RV orbit from \citet{horch2021}. For the Aa,Ab orbit, we subtract the long period motion, and we subtract the Aa,Ab short period motion to plot the long period variation for A+B. }
\label{rv_orbits2}
\end{figure}

Our ARMADA epochs are ideal for precise and quick differential astrometry of wide binaries (compared to the interferometric field-of-view). To make a large number of observations feasible, however, we do not calibrate our data with on-sky sources. Uncalibrated data, combined with short epochs, make it difficult to detect directly the inner companions seen in the ``wobble'', which are often located $\leq$1 mas (near the resolution limit of both interferometers). Still, in three cases here we are able to successfully recover the flux and position of the inner short period companions. In these cases we directly measure $a_1$ and $a_2$ of the inner orbits, which gives the inner mass ratio and inner mass sum. When we have RV data to compute a distance, we can solve for all masses independent of any other information from the literature. Without RV data, we can obtain masses by assuming a Gaia or Hipparcos distance. Figure \ref{outer_orbits2} plots the best-fit outer orbits for these three systems. In Figures \ref{period_searches2} and \ref{inner_orbits2} we show the best fit periods and inner orbits of the inner companions, one of which is a new detection entirely and the other two are first astrometric detections. In these cases, we plot the position of each inner component orbiting the center-of-mass of the system. Figure \ref{rv_orbits2} shows the RV data for HD 1976 and HD 173093, which had previously published RV orbits. We give notes on the systems in the following section.

\subsubsection{HD 1976}
HD 1976 (HIP 1921, V746 Cassiopeiae) is a B-type binary with a $\sim$170 yr orbital period for the outer visual A,B pair. This system is also a known SB1 type RV system, with \citet{blaauw1963} and \citet{abt1990} finding inner orbital periods of 27.8 and 25.44 days, respectively. \citet{mcswain2007} obtained updated RV measurements on this spectroscopic binary, and settled on a best-fit period of 25.4176 days after combining all data sets. 
\citet{harmanec2018} acquired additional velocities
and determined several spectroscopic solutions that
confirmed that the broad-lined component is the
primary of the 25 day orbit.
We acquired 5 new RVs of the broad-lined component ($v~sin~i$
= 137 km~s$^{-1}$) with the 2m AST, which are consistent 
with the 25 day SPEFO orbit of \citet[Table 3][]{harmanec2018}.
We combine the historical RV data with our new ARMADA astrometry.

We obtained 9 CHARA/MIRC-X epochs on this system, and we were able to fit for the outer binary component and detect an additional ``wobble" motion due to the inner triple. Grid searching for an inner period purely to the astrometry data, we see the strongest peak at 25.3 days, though a peak at 270 days is nearly as strong (Figure $\ref{period_searches2}$). The strength of other peaks could hint that this system has additional unseen companions, or that we have a low number of data points with astrometry alone. We find that the 270-day period does not fit the RV data at all, and there are also visual outliers in the astrometry-only long period fit. We then combine the RV and astrometry with an initial guess at the previously known inner period of 25.4 days, giving a highly inclined orbit for the inner pair. 

At the large distance of $406 \pm 54$ pc from Gaia, we expect the semi-major axis of this inner component to be on the order of $\sim$1 mas. This is difficult to detect with our uncalibrated interferometry data, since it is near the resolution limit. Still, we perform a search for flux from a third component in our interferometric phases. We are able to recover a close-in component in all of our epochs, though the astrometry errors of the Aa and Ab component are often large in this case. Figure \ref{inner_orbits2} shows our best-fit astrometry (Aa,Ab + B), using the previously published RV data to help constrain the orbit.

There is a large error in the distance measured by Gaia (406 $\pm$ 54 pc, \citealt{gaia_dr3}) and Hipparcos (307 $\pm$ 59 pc, \citealt{vanLeeuwen2007}), although the measurements agree within these large error bars. Unfortunately these distances lead to a mass sum with high uncertainty, 9 $\pm$ 5 M$_{\odot}$ for the Gaia distance. Our distance from the orbit is much lower at 186 $\pm$ 24 pc, but is too low for the masses of the B-type system. This is likely due to the noisy RV orbit biasing the semi-amplitude. RV orbits are difficult to measure for blended B-type stars when only one component is detected (e.g., \citealt{klement2021}, which shows a large discrepancy in semi-amplitude between the single-lined RV orbit from \citealt{rivinius2006} and the updated measurements with all components detected). This makes it hard to know individual masses from our data alone without more precise RV data, and the distance is not known well enough yet from Gaia. Since we detect both inner orbits, our data does tell us that the mass ratio of the Aa,Ab pair is 1.57 $\pm$ 0.28. Given the mass estimate for the primary of 6.348 M$_{\odot}$ from \citet{kervella2019}, this implies a mass for the Ab component of 4.0$\pm$0.7 M$_{\odot}$. We also compute mutual inclinations of either 59.5$\pm1.6^{\circ}$ or 130.2$\pm1.6^{\circ}$.

\subsubsection{HD 173093}
The triple system HD 173093 (HIP 91880) was recently characterized by \citet{horch2021} with spectroscopic data and speckle data at the same time as we were taking CHARA/MIRC-X data for ARMADA on the system. All three components of the triple system were detected in the spectra, allowing a full comparison of the system with speckle data to characterize the outer orbit of the visual system. The system consists of a wide A,B orbit with a 7.2 yr period, and an inner Aa,Ab orbit around the primary with a period of 2.358 days. With high precision ARMADA epochs, we can better constrain the outer orbit of this system. We are also able to uncover flux from the inner spectroscopic triple (the Ab component). This provides an independent check of the masses and inner orbit provided in \citet{horch2021}. Using the astrometry data alone, we confirm an inner ``wobble'' period of 2.358-days with a ``wobble'' semi-major axis of $\sim$0.3 mas for the primary Aa component.

We perform a joint fit to all of the RV data from \citet{horch2021}, speckle data from the WDS catalog, new ARMADA epochs measuring the outer Aa to B separation (including the inner wobble motion), and new ARMADA data measuring the inner Aa,Ab separation. Since we measure all RVs and inner/outer semi-major axes, we are able to solve for the masses of all three components, the distance to the system, and the mutual inclination between the orbits. We measure a distance to the system of 73.79$\pm$0.21 pc, which agrees well with the Hipparcos value of 73$\pm$4 pc \citep{vanLeeuwen2007}. We measure the three masses of 1.34 $\pm$ 0.06, 1.27 $\pm$ 0.06, and 1.65 $\pm$ 0.12 M$_{\odot}$ for components Aa, Ab, and B respectively. The mutual inclination for this system between the inner and outer orbits is 90 $\pm$ 4$^{\circ}$.

\subsubsection{HD 220278}
HD 220278 (97 Aqr, HIP 115404) is a bright A-type binary system with a 64.6 year orbital period and eccentricity of 0.4 \citep{malkov2012}. This orbit has been relatively well sampled with speckle data in WDS over the past decades, and hence the authors were able to compute a dynamical mass of 4.54$\pm$0.93 M$_{\odot}$. We followed the orbit with ARMADA using VLTI-GRAVITY over a previously uncovered portion of the orbital arc. Right away, we were able to detect giant residuals to our high precision data at the $\sim$10 mas level. Our period search detects a clear signal at 960 days for an inner orbital period, with a ``wobble'' semi-major of 11 mas. Given the estimated mass of $\sim$2.0-2.5 M$_{\odot}$ for the primary, we expected a semi-major axis of around 40 mas if the new component was around the primary. We search for the flux from this companion, and indeed we find a consistent orbit with the ``wobble" motion with a $K$-band flux ratio of $f1/f2$ = 20-25. The residuals from the orbit are higher than normal, hinting at either additional companions in the system or higher uncertainty due to the low flux ratio close companion.

This target does not yet have a distance in Gaia, likely due to its multiplicity. Hipparcos measures a distance of $64 \pm 3.4$ pc, giving a total mass sum of $4.5 \pm 0.7$ M$_{\odot}$. Since we measure the visual orbit of Aa,Ab we know the inner mass sum is $3.0 \pm 0.5$ M$_{\odot}$. This gives individual masses for Aa, Ab, and B components of $2.17 \pm 0.34$ M$_{\odot}$, $0.79 \pm 0.12$ M$_{\odot}$, and $1.57 \pm 0.27$ M$_{\odot}$. We measure a mutual inclination between the orbits of either 15.1$^{\circ}$$\pm$0.2$^{\circ}$ or 152.1$^{\circ}$$\pm$0.2$^{\circ}$. 

We followed this target with new spectra to measure the RV variations from the newly detected component. 
The primary is very broad with a $v \sin i =$ of 121 km~s$^{-1}$, which makes it difficult to measure 
accurately. However, we conclude that it is relatively constant in RV over the 80 days of coverage. 
This is consistent with the long inner period, and hence we will need a longer time baseline 
to measure the 960-day RV orbit. With the solar line list, a very weak, rather narrow feature 
with a $v$~sin~$i$ of 10 km~s$^{-1}$ and average line depth of $\sim$0.4\% is detectable in about 
half of our obtained spectra. We confirm that this component also shows a variable 
velocity, allowing us to conclude that there is likely a fourth companion around the B component (configuration Aa,Ab + Ba,Bb). The phase coverage is minimal for the RV orbit however, and we require more data to measure accurately the semi-amplitude and period of the new orbit. The wobble for a fourth companion is likely small, since our astrometric orbit has a median residual $\sim$100 $\mu$as. This makes it difficult to search for an additional period in the system without a higher number of epochs. Further monitoring of both RV and astrometry is needed to characterize a potential fourth body in the system.

\subsection{Systems with Wobble Only}
\label{sec:detections3}

\begin{figure}[H]
\centering
\includegraphics[width=7in]{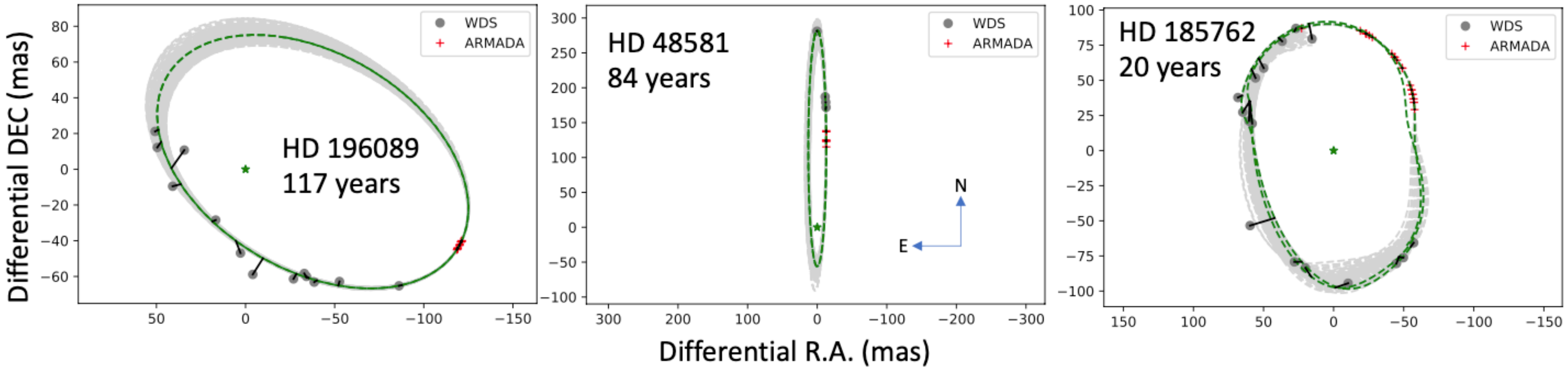}
\caption{We show our ARMADA epochs along with published WDS data for the binary systems that do not have any new RV data. The orbital coverage is quite poor for HD 196089 and HD 48581. This is especially true for HD 48581, where we present the first guess at the orbital elements. Due to its long period and high amplitude ``wobble" orbit, the shape of the outer orbit is visually distorted for HD 185762. We show the best-fit orbit for two revolutions of the orbital period in order to match up in time with the early WDS data and our recent ARMADA epochs. The grey shaded regions depict 100 random orbits from the MCMC chains, for one orbital revolution. }
\label{outer_orbits3}
\end{figure}

\begin{figure}[H]
\centering
\includegraphics[width=7in]{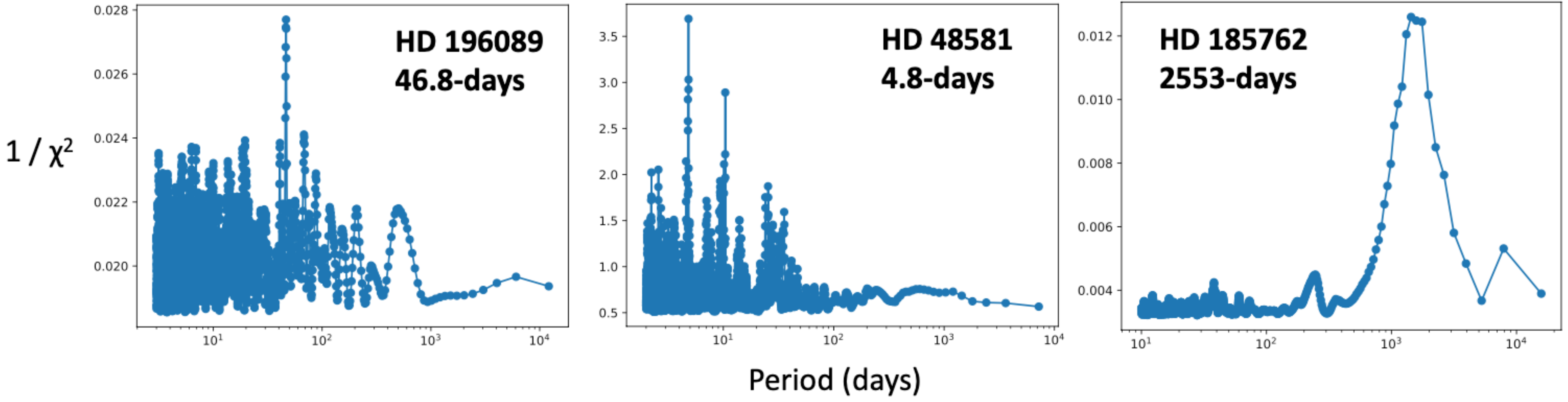}
\caption{Our residuals to the pure binary fit are high for the systems in Figure \ref{outer_orbits3}, and we see a clear peak in the period search for inner companions. In HD 196089 and HD 48581, the ``wobble" amount is relatively small at $<$200 $\mu$as. For HD 185762, our best fit wobble is $>$10 mas and the period search is aided by including historical data from WDS for the long period ``wobble" motion. }
\label{period_searches3}
\end{figure}

\begin{figure}[h]
\centering
\includegraphics[width=4.5in]{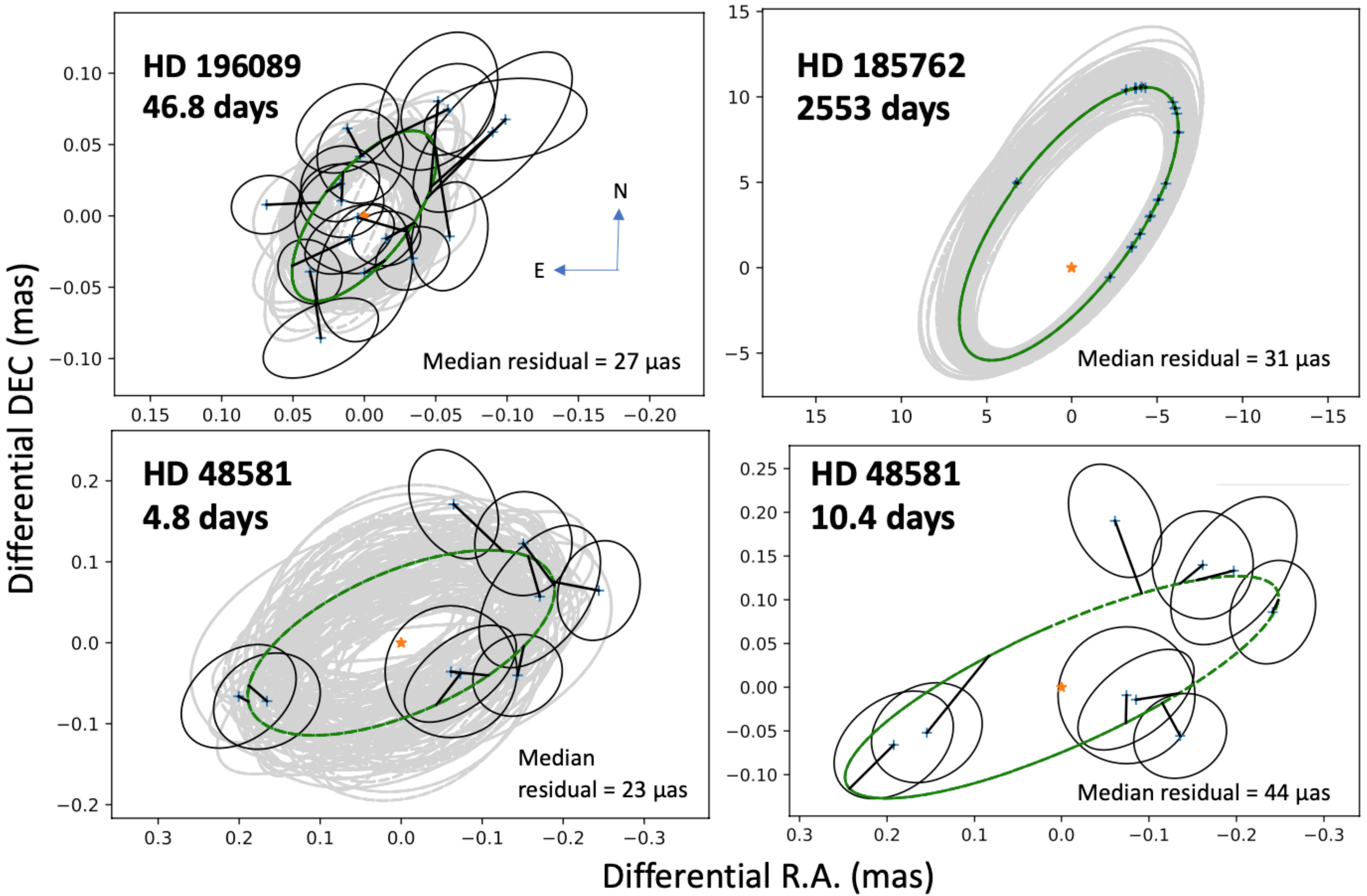}
\caption{We plot the best-fit inner orbits due to the newly detected companions from wobble motion, after subtracting out the outer binary motion shown in Figure \ref{outer_orbits3}. For HD 48581 we plot the two convincing period peaks in the periodogram, though the solution at 4.8 days is preferred. We see a median residual $\leq$30 $\mu$as in these newly detected systems, though further information is required to solve for individual masses in these cases. This can be accomplished with RV monitoring, a better outer orbit, or a better distance from future Gaia releases. The grey shaded regions show that inner orbit errors are higher in cases which do not have any RV data or flux detected from the inner companion.  }
\label{inner_orbits3}
\end{figure}

For three of our new detections, we were unable to obtain RV orbits. However, non-detections in the RV can help us to solve which component hosts a companion, as described below. With our pure astrometry orbit, we can detect new companions and constrain the orbital period and elements. We can also obtain a total mass sum, although two of these outer binaries happen to be poorly constrained. This leads to large errors in the mass sum, and makes it difficult to further constrain individual masses at this point. We show the outer binary orbits in Figure \ref{outer_orbits3}, and the search for additional orbiting companions in Figure \ref{period_searches3}. Figure \ref{inner_orbits3} shows the best-fit inner ``wobbles" with the binary motion subtracted. In these ``wobble"-only cases there is a higher uncertainty on the inner orbit, which can especially be seen in the MCMC orbits for the small wobbles of HD196088/9 and HD48581. This is because the orientation of the inner orbit depends on all 14 orbital parameters (subtraction of the outer orbit, and wobble of the inner orbit), and there are no additional RV data or flux detected from the new companion to constrain the orientation. However, in all cases the median residual to the orbit fit is quite good, though further RV/orbit monitoring is required to solve for masses. We give notes on each system in this section.

\subsubsection{HD 196088/9}
HD 196088/9 (HIP 101398) is a close visual binary with a composite spectrum consisting of B9.5~V and G9.5~III components. This A,B pair has an orbital period of 185 yr \citep{hartkopf2009}. There is a much wider tertiary component (designated component C in WDS) associated with the system at 4 arcsecond separation. However, with only an arc of the orbit covered, dynamical masses are uncertain for the A,B system. We follow this long period binary with CHARA-MIRC-X, covering a new portion of its orbit to improve the outer orbital elements. In addition, we discover a strong period signal at 50 days that hints at a third companion orbiting one of the stars in the system. The wobble is relatively small at $<$100 $\mu$as, but given the high mass of the system and its distance, the new companion is still of stellar mass. We fit a circular orbit for the new ``wobble" motion since $\omega$ and $T_0$ become degenerate for near-circular orbits, especially when we do not have additional RV to constrain the orbit.

We attempted to follow-up this detection with new RV data. We found that the spectrum is 
dominated by the slowly rotating late-type giant, which has a $v$~sin~$i$ of 6 km~s$^{-1}$. 
With the use of the normally used A star line list,
lines of the early-type component are not obvious. Thus, we next tried using a line list 
between 4400--4900~\AA, a spectral region blueward of our usual A star line
list. In that wavelength region there is possible evidence of the early-type star, but 
if so, its lines are broad and very weak, and so, unmeasurable given the dominance of the
late-type giant. For the late-type giant we determined an average RV of $-$28.4 $\pm$ 0.1 
km~s$^{-1}$ over a 230-day observing period indicating that its velocity is constant
over that interval. This means that the new tertiary must be around the early-type component. 
This component is the brighter primary of the binary system, making its configuration Aa,Ab + B. 

Since we do not yet have RV data on the newly discovered inner orbit, we need to rely on distance measurements to compute mass information. Hipparcos measured a distance of 460 $\pm$ 60 pc \citep{vanLeeuwen2007}, while Gaia measured 530 $\pm$ 140 pc \citep{gaia_dr3}. Using the more precise distance measurement from Hipparcos, we compute a total mass sum of 8 $\pm$ 3 M$_{\odot}$ for the system. Without a more reliable distance estimate, we cannot constrain individual masses to good precision. With future RV data, or a better estimate from upcoming Gaia releases, we will be able to constrain the parameters of this system to a better degree. We find that the mutual inclination of the orbits is 77 $\pm$ 8$^{\circ}$ or 83 $\pm$ 8$^{\circ}$. In either case, the inclinations between the inner and outer orbits are not well aligned.

\subsubsection{HD 48581}
HD 48581 (HIP 32140) is an early F-type binary, only sparsely covered in WDS with three recent speckle points from 2015--2017 and one Hipparcos data point from 1991. Our ARMADA data from VLTI-GRAVITY is not very far from these epochs in terms of orbital fraction of the long period, making the binary orbit very poorly constrained. We offer first estimates of the outer orbital elements, though we acknowledge that our high eccentricity solution is highly uncertain and likely to be updated with the addition of new data in the future. Still, we are able to search for additional short period ``wobble'' motions which are not heavily affected by the long period binary orbit. We find a promising detection of a third companion in the system, with a period of 4.8 days and wobble semi-major of 0.2 mas, although there are other peaks in the periodogram shown in Figure \ref{period_searches3}. We report the two best solutions in Table \ref{table:elements}, shown in Figure \ref{inner_orbits3}. Both solutions are near-circular, and we report the orbital elements from a circular fit to the inner orbit. Additional astrometric or RV monitoring is needed to confirm this inner period.

We attempted to follow this system with the 2m AST, but the southern declination of 
$-32.7^{\circ}$ resulted in reduced signal-to-noise ratios for the spectra and 
an enhancement of water vapor lines. The spectra are dominated by the very broad lines,
having a $v$~sin~$i$ of 120 km~s$^{-1}$, of the primary component. However, 
most spectra also have one or more very weak relatively narrow features that on average 
are less than 0.5\% deep and are positioned inside the broad lines. 
The RV of the broad-lined primary ranges from 9.6--14.4
km~s$^{-1}$, but so far we cannot detect evidence 
of an orbital period that fits the astrometric variation.
The RV range may simply result from measurement
uncertainties of the very broad lines. We are also 
unable to fit the apparent RV variability 
of the weak features with the 4.8 day astrometric period.
These features instead may result from pulsation and at times
from water vapor lines. Since the primary does not seem to show 
large variation in its RV, this makes it likely that it is the wide 
B component that is itself a short period binary (A + Ba,Bb). Additional 
RV data are needed to confirm this configuration.

Since we do not yet have RV data on the newly detected inner orbit, we are not able to compute an independent measure of distance. Gaia measured a distance of 198.3 $\pm$ 3.1 pc \citep{gaia_dr3}, while Hipparcos has less precision on the distance of 188 $\pm$ 29 pc \citep{vanLeeuwen2007}. Though the outer orbit is quite unreliable with only a small arc of its long period covered, we use the Gaia distance to compute a total mass sum of 5.3$\pm$1.2 M$_{\odot}$. Unfortunately the distance to the system is likely too high to resolve the inner component with interferometry (even when well calibrated), since the companion is expected to be $<$1 mas in semi-major axis. We are able to measure the mutual inclination of the system to be either 68$\pm$16$^{\circ}$ or 118$\pm$16$^{\circ}$.

\subsubsection{HD 185762}
HD 185762 (45 Aql, HIP 96807) is a known visual binary, with a primary spectral type ranging from A0 -- A3 in Simbad. \citet{hartkopf2000} published the first orbit for the visual binary with a 21-year orbital period. \citet{mason2010} updated this orbit, finding a period of 20.3 years and a mass sum of 1.87$\pm$0.59 M$_{\odot}$. These authors pointed out that the mass sum was likely much too low for the spectral type of the primary, calling into question the parallax of the system measured by Hipparcos. \citet{malkov2012} most recently characterized the visual orbit of this system, also finding a 20.3-year orbital period for the binary. They found the same inconsistency between the dynamical mass of 1.88$\pm$0.45 M$_{\odot}$ and the photometric mass of 3.34 M$_{\odot}$. This either implies an incorrect outer orbit, or a distance that is too low. 

We followed this system for three years with CHARA/MIRC-X. Our high precision astrometric data did not fit well with the historical data from WDS, motivating a search for additional companions. We find a clear long-period peak in the system at $\sim$2000-days which fits our ARMADA arc of data well and also is consistent with WDS (Figures \ref{outer_orbits3} and \ref{period_searches3}). The orbit leads to an inner wobble of amplitude $\sim$11 mas (with only 31$\mu$as median residuals), though we are unable to detect flux from a third component in the system. With a lack of RV data for this target, we cannot deduce which component of the binary the new companion is orbiting. Hipparcos and Gaia measure a distance of 103.4$\pm$1.7 pc and 108$\pm$8 pc, respectively \citep{vanLeeuwen2007,gaia_dr3}. Taking the higher precision Hipparcos measurement, we compute a dynamical mass of 1.93$\pm$0.13 M$_{\odot}$. This value agrees with previous studies, but is still too low for the spectral type of the system. Since the outer orbit is now relatively well covered, this implies that the distance to the system is likely too low. Our newly detected companion with a large ``wobble" motion is likely biasing the parallax measurements of Hipparcos and Gaia. It is possible that the distance will be improved in future Gaia releases. However, the outer binary is at the inner edge of Gaia's field-of-view, which could make the modeling for a three-body system more difficult. A better distance measurement is needed to solve for individual masses of this system. 

Our newly detected inner orbital period of 7-years is quite large for an outer binary period of only 20-years. The outer semi-major axis of 90 mas is only $\sim 9 \times$ larger than the ``wobble" semi-major of 11 mas. At a mass ratio of 1:1, the minimum possible inner semi-major axis is 22 mas ($\sim$1/4 of the outer binary semi-major axis). However, we are unable to recover the flux from the newly detected component. This implies that the semi-major is larger than 22 mas. Even at a mass ratio of 3:1, the inner semi-major axis becomes 44 mas which is nearly half of the semi-major of the outer orbit. A well calibrated interferometric epoch is needed to detect the flux from this companion, which seems to push the limits of orbital stability in triples. We compute a mutual inclination of either 51.6$\pm$1.8$^{\circ}$ or 96.2$\pm$1.9$^{\circ}$, meaning that the orbits are not well-aligned in either case. 

This system is particularly interesting for follow-up study. First of all, its dynamical mass is inconsistent with its spectral type (especially since there are now three components within the system). Orbital stability is also an issue for the newly detected companion, with the outer/inner orbital periods having a ratio very close to 3:1. This allows for the possibility of resonant behavior, which could explain the stability of the system. It must be noted that a period ratio of 3 does not guarantee resonance, as such a ratio is necessary but not sufficient. 

\begin{longtable}{lllr}
\caption{System Information}
\label{table:masses}\\
\hline
HD (configuration) & Flux ratio (f1/f2)\tablenotemark{a} & Mass (M$_{\odot}$)\tablenotemark{b}  & Mutual Inclination ($^\circ$)\tablenotemark{c} \\
\hline
\endhead

199766 (Aa,Ab+B) & 1.8$\pm$0.3 (H, A/B) & M$_{dyn}$ = 4.7$\pm$0.9 & 52.2$\pm$2.5 / 126.8$\pm$2.5 \\
 & 1.62$\pm$0.06 (K, A/B) & M$_{Aa}$ = 2.7$\pm$0.9 & \\
 & & M$_{Ab}$ = 0.45$\pm$0.09 & \\
 & & \textit{M$_B$ = 1.59} & \\
 
\hline

29573 (A+Ba,Bb) & 4.1$\pm$0.3 (H, A/B) & M$_{dyn}$ = 4.2$\pm$0.5 & 9.0$\pm$0.7 / 160.2$\pm$0.7 \\
 & 4.1$\pm$0.4 (K, A/B) & \textit{M$_{A}$ = 2.65} & \\
 & & M$_{Ba}$ = 1.0$\pm$0.4 & \\
 & & M$_{Bb}$ = 0.60$\pm$0.15 & \\
 
\hline
\newpage

31297 (A+Ba,Bb) & 1.4$\pm$0.2 (K, A/B) & M$_{dyn}$ = 3.8$\pm$1.0  & 58.6$\pm$0.4 / 63.3$\pm$2.8 \\
 & & \textit{M$_{A}$ = 1.4} & \\
 & & M$_{Ba}$ = 1.7$\pm$0.7 & \\
 & & M$_{Bb}$ = 0.74$\pm$0.27 & \\

\hline

37711 (A+Ba,Bb) & 4.6$\pm$0.5 (H, A/B) & M$_{dyn}$ = 4.6$\pm$2.3 &  60$\pm$14 / 82$\pm$32 \\

\hline

5143 (Aa,Ab+Ba,Bb) & 2.5$\pm$0.3 (K, A/B) & M$_{dyn}$ = 4.05$\pm$0.23  & 80$\pm$30 / 120$\pm$30 (Aa,Ab)  \\
 & & & 47.2$\pm$2.0 / 101.6$\pm$2.0 (Ba,Bb) \\

\hline

16753 (A+Ba,Bb) & 2.04$\pm$0.07 (K, A/B) & M$_{dyn}$ = 4.7$\pm$0.5  & 20$\pm$4 / 50$\pm$4 \\
 & & \textit{M$_{A}$ = 2.063} & \\
 & & M$_{Ba}$ = 1.9$\pm$0.4 & \\
 & & M$_{Bb}$ = 0.68$\pm$0.09 & \\

\hline

1976 (Aa,Ab+B) & 3.2$\pm$0.7 (H, Aa/B) & M$_{dyn}$ = 9$\pm$5 & 59.5$\pm$1.6 / 130.2$\pm$1.6 \\
 & 12$\pm$3 (H, Aa/Ab) & \textit{M$_{Aa}$ = 6.348} & \\
 & & M$_{Ab}$ = 4.0$\pm$0.7 & \\

\hline

173093 (Aa,Ab+B) & 1.1$\pm$0.1 (H, Aa/B) & M$_{dyn}$ = 4.3$\pm$0.1 & 90$\pm$4 \\
 & 1.36$\pm$0.06 (H, Aa/Ab) & M$_{Aa}$ = 1.34$\pm$0.06 & \\
 & & M$_{Ab}$ = 1.27$\pm$0.06 & \\
 & & M$_{B}$ = 1.65$\pm$0.12 & \\

\hline

220278 (Aa,Ab+B) & 3.0$\pm$0.3 (K, Aa/B) & M$_{dyn}$ = 4.5$\pm$0.7 & 15.1$\pm$0.2 / 152.1$\pm$0.2 \\
 & 21$\pm$4 (K, Aa/Ab) & M$_{Aa}$ = 2.17$\pm$0.34 & \\
 & & M$_{Ab}$ = 0.79$\pm$0.12 & \\
 & & M$_{B}$ = 1.57$\pm$0.27 & \\

\hline

196088/9 (Aa,Ab+B) & 15$\pm$1 (H, A/B) & M$_{dyn}$ = 8$\pm$3 & 77$\pm$8 / 83$\pm$8 \\

\hline

48581 (A+Ba,Bb) & 5.4$\pm$0.3 (K, A/B) & M$_{dyn}$ = 5.3$\pm$1.2  & 68$\pm$16 / 118$\pm$16 \\

\hline

185762 (unknown) & 4.1$\pm$0.4 (H, A/B) & M$_{dyn}$ = 1.93$\pm$0.13  & 51.6$\pm$1.9 / 96.2$\pm$1.9 \\

\hline

\end{longtable}
\tablenotetext{a}{Mean and Standard Deviation of all fitted values}
\tablenotetext{b}{Italics = Mass assumed as described in text}
\tablenotetext{c}{Table shows two possible values for most systems}

\section{Mutual Inclinations for New Detections}
\label{sec:inclinations}
\begin{figure}[H]
\centering
\includegraphics[width=4.5in]{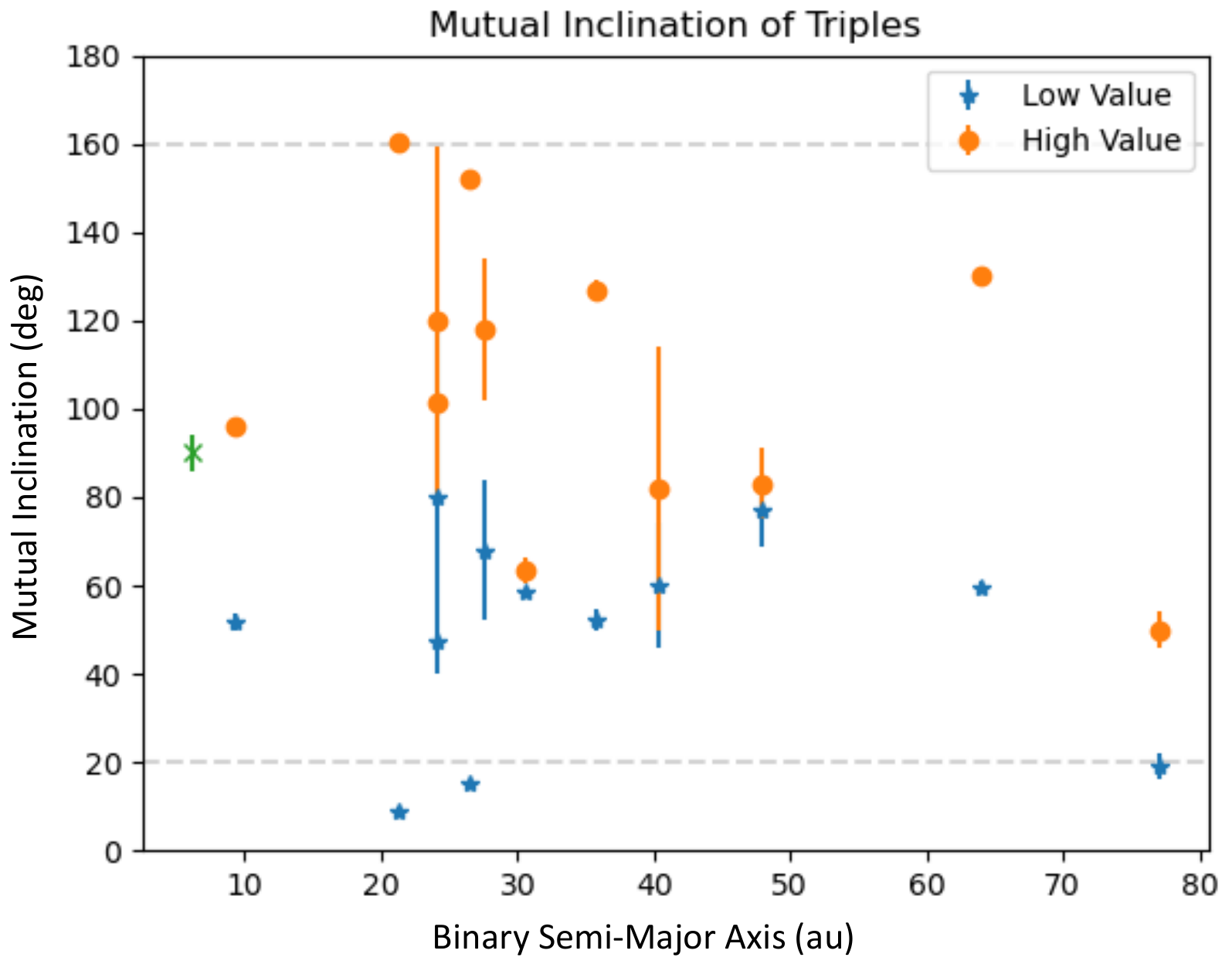}
\caption{We plot the mutual inclinations between the inner and outer orbits of our systems. Without RV information on the long period outer binary orbit, most of our systems have two possible measurements for mutual inclination. We show both possibilities, split between the lower inclination and higher inclination options in the plot. The degeneracy does not exist for HD 173093 (green x in the plot). The prevalence of misaligned orbits is somewhat surprising, as \citet{tokovinin2017} found that triples with an outer binary semi-major axis $<$50 au are generally aligned within 20$^{\circ}$ (depicted in the prograde and retrograde case with grey dashed lines). Only three of our systems potentially fall within $<$20$^{\circ}$ of the outer orbit orientation, hinting that higher mass triples have more misalignments. }
\label{mutual_inclinations}
\end{figure}

Most of our newly detected inner systems are significantly misaligned with their outer binary orbit. We show the dependence of mutual inclination on the outer binary separation in Figure \ref{mutual_inclinations}. To compute the binary semi-major axis, we use the orbital periods in Table \ref{table:elements} and dynamical masses in Table \ref{table:masses}. For most of our systems there are two mutual inclination solutions possible, and we cannot break the degeneracy since we are missing RV information on the outer binary orbit. This could be solved in future Gaia releases, when 5-year monitoring of RV orbits is published. In any case, it is apparent that very few of our systems have a mutual inclination $<$20$^{\circ}$ (only three potentially fall within this range of the outer orbit). This is somewhat surprising, as \citet{tokovinin2017} found that triples with the outer binary $a<50$ au are generally well aligned (mutual inclination $<$ 20$^{\circ}$), and it is not until $a>1000$ au where misalignments become more common. However, that work also hinted that higher mass stars in general lead to more misalignments in triple systems. \citet{borkovits2016} also found a preferential alignment of triples with their Kepler sample (and hence a lower mass sample than ours), though they detected a peak at 40$^{\circ}$ alignment which is likely due to Kozai-Lidov cycles. That sample also included more compact triples than the ones presented here, and it is not obvious that we would see a similar peak for wider triple systems. \citet{dupuy2022} studied alignments of Kepler planets in binary systems and likewise found a preference for alignment between the orbits, although the formation mechanisms for planets are different than for stellar triples.

Most studies for mutual inclination of triples have focused on solar-type and lower mass stars. More massive stars are thought to be prone to dynamical interactions with their companions since they often form within dense environments. For example, companions attained through capture processes are likely to be misaligned. In addition, companion formation via disk instability and/or rapid inward migration can lead to ejections, misaligned orbits, or eccentric orbits for inner subsystems \citep{tokovinin2021}. Our results are consistent with this general picture, as most of our early-F/A/B-type triples have misaligned orbits. The only system where we break the degeneracy of 2 possible solutions here is HD 173093, where we measure the two orbits to be perpendicular at 90$\pm$4$^{\circ}$ for a binary separation $<$10 au. Nonetheless, the formation of triple systems is a rich dynamical problem and further work is necessary to explore all of the possible mechanisms. Crucially, we must state that our ARMADA triples presented here are selected from a sample that may increase our chances of detecting misaligned systems. As will be fully described in future papers, we have selected binary systems that are wide (for the purpose of searching for companions around individual components of the wide pair) but that are also within the field-of-view of the interferometer (so we can carry out our differential astrometric methods). This biases us towards visual binaries with wide physical separations but low projected separations, which occurs in inclined systems. Moreover, it is reasonable to expect astrometric ``wobbles" to be more easily detected when they are viewed face-on. This combination potentially biases us toward detecting systems with high mutual inclinations. However, we point out that we have detected many inclined ``wobbles" in this paper. Though we see some hint for high mutual inclinations for intermediate mass triple systems, a more thorough statistical study of the full sample and its biases are needed once data collection for ARMADA is complete.

\section{Conclusions and Future Work}
\label{sec:conclusion}

Our ARMADA survey with the MIRC-X instrument at the CHARA Array and the GRAVITY instrument at VLTI is currently underway to probe for companions down to the giant planet regime on $\sim$au orbits around individual stars of binary systems. Confident detections of giant planets require a long time-baseline and a higher number of epochs, but with our high precision we can more easily see the ``wobble'' signature from the gravitational tug of previously unseen short-period tertiary companions. In this paper, we publish astrometric orbits for 9 newly detected inner components to known binaries. We also publish first astrometric orbits for 3 previously known triple systems. For 8 of these systems we combine our astrometry with RV data to confirm the newly detected periods and solve which component of the binary the new companion is orbiting. We publish first RV orbits for 6 of these systems. We see a preference for misaligned systems in the triples discovered so far in the ARMADA survey. Though this picture is consistent with dynamical reprocessing leading to more misalignments in intermediate mass triples, we note that we have not yet accounted for potential biases in target selection.

In addition to discovering new triple systems, we have three sources here that have both GRAVITY and MIRC-X data. For MIRC-X, we use an etalon calibration system to ensure that our astrometry is internally consistent \citep{gardner2021}. GRAVITY data on the other hand are on an absolute wavelength scale which is accurate at the $5 \times 10^{-5}$ level \citep{sanchez2017}. For shared sources, we can fit a scale factor between the datasets to bring all MIRC-X nights with etalon data to the same absolute scale as GRAVITY. We compute a scale factor of $1.00495\pm0.00017$, $1.00501\pm0.00021$, and $1.0061\pm0.0005$ for systems HD 199766, HD 29573, and HD 31297 respectively. Averaging the measurements gives a scale factor of $1.00535\pm0.00019$ between the binary separations measured by GRAVITY and MIRC-X. If etalon data is taken on a MIRC-X night, this factor can be applied to bring the data to an absolute wavelength scale (e.g., for binary separations on nights brought to the same internal astrometric scale with the etalon, MIRC-X separations should be divided by this value). Otherwise there is a 0.5\% systematic error, which in the case of binary stars would be applied to the measured separation of the binary with MIRC-X. Note that this is higher than the 0.25\% systematic estimated for the older MIRC instrument \citep{monnier2012}. 

For these multiple systems, our astrometric precision is regularly at the 20-50 $\mu$as level when performing a joint fit with RV (though there are some outliers, possibly due to yet unidentified companions). This is promising astrometric precision for probing the brown dwarf and giant planet mass regime in binary systems. We are currently following $\sim$70 binary systems with the ARMADA survey, and future papers will analyze our detection limits around all stars in the system to constrain the giant planet / brown dwarf / stellar mass frequency for the $\sim$au regime around intermediate mass stars. Our new detections published here will be useful for studying the inner triple rate of intermediate mass binaries, once ARMADA is complete. This is a measurement that is difficult to obtain with other methods, given the sparse and noisy RV information on such systems and the fact that these binaries are below the resolution limit of single dish telescopes. Future ARMADA results will also be used to study our systematics and what might be limiting $\leq$10 $\mu$as astrometry, which is needed to probe down to 1 Jupiter mass or lower for these systems. 

\acknowledgments
T.G. and J.D.M. acknowledge support from NASA-NNX16AD43G and from NSF-AST2009489. T.G. acknowledges support from Michigan Space Grant Consortium, NASA grant NNX15AJ20H. Astronomy at Tennessee State
University is supported by the state of Tennessee through its Centers of Excellence program.
This work is based upon observations obtained with the Georgia State University Center for High Angular Resolution Astronomy Array at Mount Wilson Observatory.  The CHARA Array is supported by the National Science Foundation under Grant No. AST-1636624 and AST-2034336.  Institutional support has been provided from the GSU College of Arts and Sciences and the GSU Office of the Vice President for Research and Economic Development. MIRC-X received funding from the European Research Council (ERC) under the European Union's Horizon 2020 research and innovation programme (Grant No. 639889). JDM acknowledges funding for the development of MIRC-X (NASA-XRP NNX16AD43G, NSF-AST 1909165).
S.K. acknowledges support from ERC Consolidator Grant (Grant Agreement ID 101003096) and STFC Consolidated Grant (ST/V000721/1).
The research leading to these results  has received funding from the European Research Council (ERC) under the European Union's Horizon 2020 research and innovation program (project UniverScale, grant agreement 951549).
This research has made use of the Jean-Marie Mariotti Center SearchCal service\footnote{available at \url{http://www.jmmc.fr/searchcal_page.htm}}. This research has made use of the Jean-Marie Mariotti Center \texttt{Aspro}
service \footnote{Available at \url{http://www.jmmc.fr/aspro}}. We thank Nuria Calvet for supporting funds in the development of our etalon wavelength calibration module. This research has made use of the Washington Double Star Catalog maintained at the U.S. Naval Observatory.
Based on observations collected at the European Southern Observatory under ESO large programme 1103.C-0477.
This work has made use of data from the European Space Agency (ESA) mission
{\it Gaia} (\url{https://www.cosmos.esa.int/gaia}), processed by the {\it Gaia}
Data Processing and Analysis Consortium (DPAC,
\url{https://www.cosmos.esa.int/web/gaia/dpac/consortium}). Funding for the DPAC
has been provided by national institutions, in particular the institutions
participating in the {\it Gaia} Multilateral Agreement.

%
%



\vspace{5mm}
\facilities{CHARA, VLTI, Fairborn Observatory}
\software{lmfit, emcee}



\bibliographystyle{aasjournal}
\bibliography{references}

\listofchanges

\end{document}